\documentclass[lettersize,journal]{IEEEtran}
\usepackage{amsmath,amsfonts}
\usepackage{algorithmic}
\usepackage{algorithm}
\usepackage{array}
\usepackage[caption=false,font=normalsize,labelfont=sf,textfont=sf]{subfig}
\usepackage{textcomp}
\usepackage{stfloats}
\usepackage{capt-of}
\usepackage{url}
\usepackage{verbatim}
\usepackage{graphicx}
\usepackage{cite}
\usepackage{xcolor}
\usepackage{booktabs}

\newtheorem{proposition}{Proposition}

\begin{document}

	\title{Assignment-Free Swarm Navigation via Adaptive Target-Field Planning and Mean-Shift Exploration}

	\author{Jiaxuan Li, Junhui Zhang,~\IEEEmembership{Member,~IEEE,} and Bo Zhu,~\IEEEmembership{Member,~IEEE}
		\thanks{This work was supported by the National Natural Science Foundation of China under Grant 62603330. \emph{(Corresponding author:Junhui Zhang)}}
		\thanks{Jiaxuan Li, Junhui Zhang and Bo Zhu are with the School of Robotics and Automation at Nanjing University, Suzhou, Jiangsu 215163, China (e-mail: jiaxuanli@smail.nju.edu.cn; junhuiz@nju.edu.cn; zhubo@nju.edu.cn).}}



	\maketitle

	\begin{abstract}
		Mean-shift exploration enables assignment-free swarm shape assembly, but obstacle-constrained navigation additionally requires coordinated adaptation of the desired morphology and its physical execution. This paper presents an environment-adaptive navigation framework that integrates target-field planning, execution regulation, and morphology-preserving mean-shift exploration. A feasibility-aware graph search plans the position, orientation, and morphology of an unlabeled target field without specifying robot-specific goals. An execution-regulated reference manager adjusts field progression and morphology switching using swarm entering, coverage, and lag metrics. A moving-frame controller separates prescribed rigid motion from local mean-shift redistribution and incorporates neighbor interactions and obstacle responses. Simulations and hardware experiments demonstrate that the proposed framework coordinates morphology transitions and collective motion for assignment-free swarm navigation through cluttered environments.
	\end{abstract}

	\begingroup
	\renewcommand{\abstractname}{Note to Practitioners}
	\begin{abstract}
		This work addresses robot team navigation through restricted spaces while retaining an organized group shape. Such behavior could support coordinated inspection in cluttered industrial facilities. The operator provides an obstacle map, a destination, and a set of candidate group shapes, rather than individual destinations for every robot. The planner selects the route, group orientation, and shape changes according to the available space. During execution, a shared motion reference can slow down or wait when the measured robot distribution indicates that more time is needed to regroup. This removes the need to prescribe switching locations manually and coordinates shape changes with the actual progress of the team. Implementation requires robot position feedback and low-level velocity tracking. Simulations and a three-robot experiment demonstrate passage traversal and shape recovery. When adapting the method to another platform, users should select shape templates, reference speeds, and clearance margins that match robot size, sensing accuracy, and motion limits. Current evaluation covers known maps and static obstacles. Operation with moving obstacles or uncertain measurements calls for further development and safety validation.
	\end{abstract}
	\endgroup


	\section{Introduction}
	\IEEEPARstart{R}{obot} swarms provide scalable collective autonomy through local sensing, limited communication, and distributed interactions \cite{dorigo2021past}. Experiments demonstrate thousand-robot self-assembly \cite{rubenstein2014programmable} and autonomous aerial swarm navigation in natural environments \cite{zhou2022wild}. Reconfigurable physical connections support outdoor adaptation \cite{zhao2024snail} and closed-loop assembly and navigation of magnetic modular millibots \cite{bhattacharjee2026millibots}. Cyborg-insect swarms also navigate unknown obstructed terrain \cite{bai2025cyborg}. These developments motivate navigation methods that coordinate collective motion with environmental adaptation. For independently mobile robots, this includes adapting the desired morphology to surrounding free space.

	\begin{figure*}[t]
		\centering
		\makebox[\textwidth][c]{%
			\includegraphics[
			width=0.98\textwidth
			]{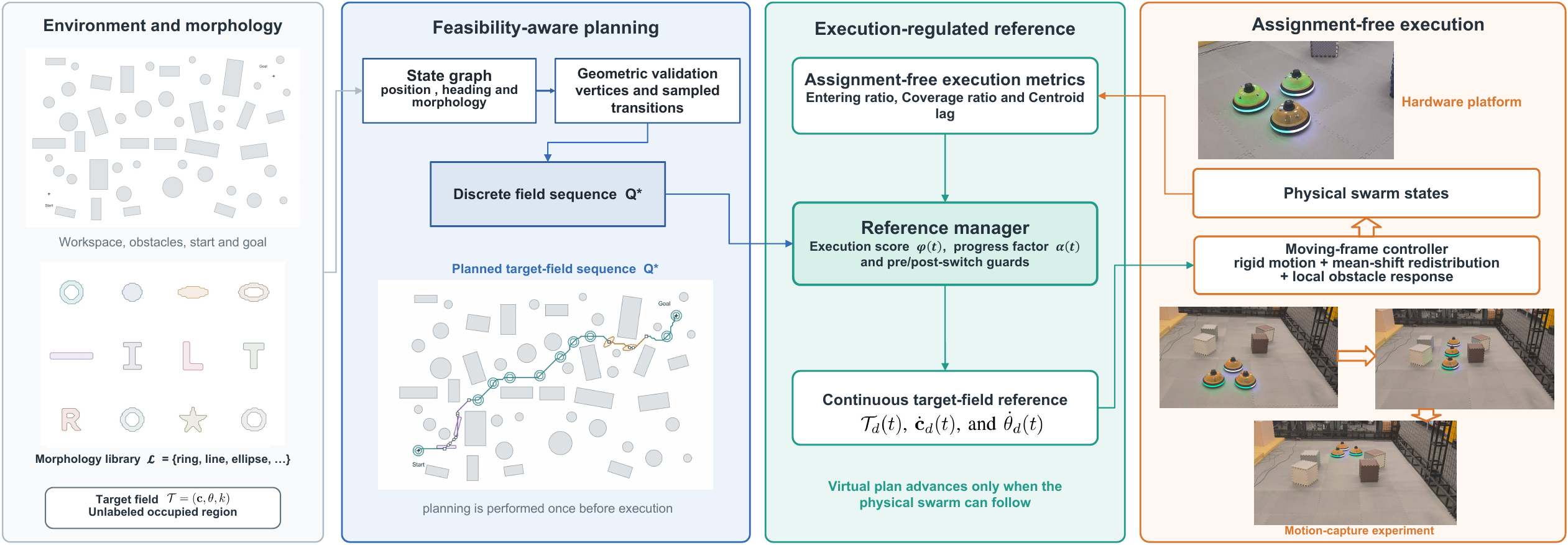}%
		}
		\caption{Conceptual overview of the proposed planning-to-execution framework. Given the workspace, obstacles, start and goal regions, and a morphology library, the feasibility-aware planner constructs a state graph over the target-field position, heading, and morphology. Geometrically feasible vertices and sampled transitions are retained to obtain a discrete target-field sequence $Q^\star$ before execution. During execution, the reference manager converts $Q^\star$ into a continuous target-field reference $\mathcal{T}_{d}(t)$ and regulates its progression using assignment-free execution metrics together with pre- and post-switch guards. The moving-frame controller realizes rigid motion, mean-shift redistribution, and local obstacle responses without persistent robot-to-target assignments, while the measured physical swarm states close the execution-regulation loop. Representative motion-capture experiments are shown on the right.}
		\label{fig:framework_overview}
	\end{figure*}

	Collective shape formation provides a basis for such geometric adaptation. Local task swapping coordinates target assignment and motion through distributed interactions \cite{wang2020shape}. CAT-ORA integrates assignment with collision-aware, time-optimal formation reshaping \cite{kratky2025catora}. Model predictive and receding-horizon control also address constrained formation reconfiguration \cite{park2023reconfiguration,zhou2025receding}. These formulations specify target positions or robot-specific relative poses. Region-occupation tasks can instead use spatial references without persistent robot-to-goal correspondence.

	Assignment-free shape assembly offers an alternative to explicit robot-to-goal assignment. Region-based control describes collective geometry without fixed robot identities \cite{cheah2009region}. Mean shift is a density-based mode-seeking procedure \cite{comaniciu2002meanshift}. Sun \emph{et al.} developed mean-shift exploration for assignment-free shape assembly using local entry, exploration of unoccupied regions, and inter-robot interactions \cite{sun2023meanshift}. Their framework also supports shape-preserving maneuvers through distributed negotiation of the desired shape reference. Extensions address precise shape formation and population-dependent shape sizing \cite{zhang2024meanshift,li2024adaptive}. Related work addresses relative localization \cite{lu2025localization} and memory-efficient two- and three-dimensional shape representations \cite{yue2026memory}. Density-driven control achieves shape formation without communication or identity recognition \cite{xiang2026density}. Continuous Sculpting maintains persistent shapes through robot circulation and adapts to user-initiated shape changes \cite{curtis2025sculpting}. These advances establish that local redistribution can sustain and modify collective shapes.

	Obstacle-constrained navigation additionally requires coordination between morphology selection and physical execution. A compact morphology may become infeasible near a narrow passage, requiring an elongated morphology to be selected and rotated before entry. After traversal, the preferred morphology may need to be recovered. Microswarm navigation further highlights the importance of incorporating actuation constraints into planning and control \cite{chu2026actuation}. For independently mobile robots, redistribution, neighbor interactions, obstacle responses, and actuation limits can delay tracking of a moving reference. Navigation therefore requires both a feasible morphology sequence and progression coordinated with swarm execution.

	Formation control offers established coordination tools \cite{oh2015formation}, including affine transformations \cite{zhao2018affine} and self-reconfigurable hierarchies \cite{zhang2024hierarchical}. Recent controllers address collision avoidance and communication maintenance \cite{liu2026modular}, as well as affine maneuvers under unreliable networks \cite{zhang2026reliable}. At the shape level, Li \emph{et al.} combine distributed clustering, constrained shape-parameter optimization, and mean-shift control to form multiple shapes \cite{li2025multiple}. These results demonstrate existing links between shape-level decisions and local robot execution.

	Formation navigation research addresses environmental constraints through complementary representations. Constrained optimization supports formation reconfiguration and cooperative transport in dynamic environments \cite{alonsomora2017formation}. Configuration-graph planning and adaptive formation scaling support passage through constrained environments \cite{liu2024graph,su2022choke}. Distributed formation planning addresses dense-environment flight, computational efficiency, and abnormal agents \cite{quan2023formation,zhou2025reformation}. DEFORM selects passage-dependent formation patterns and combines assignment with distributed model predictive control \cite{li2025deform}. Learning-based navigation also adapts formations, including under communication constraints \cite{deng2025adaptation,yao2025communication}. DSPH combines smoothed particle hydrodynamics with gradient velocity obstacles to coordinate swarm interactions and dynamic obstacle responses \cite{zhang2026dsph}.

	Execution methods further address computation, safety, and tracking performance during coordination. Policy learning accelerates distributed model predictive control \cite{zhang2025scalable}. Motion-aware control barrier functions regulate pairwise safety constraints in flocking \cite{su2026safety}. Hierarchical reinforcement learning also couples leader--follower control with prescribed performance \cite{tan2025safe}. These studies highlight computational, safety, and tracking requirements that must accompany geometric adaptation.

	At the microswarm scale, magnetic actuation enables pattern adaptation for channel navigation \cite{yu2022magnetic} and learned actuation parameters for elliptical shape formation \cite{park2025elliptical}. Deep learning supports environment-adaptive distribution planning \cite{yang2022distribution}, while reinforcement learning enables navigation under partial observation in unknown environments \cite{an2026unknown}. These studies demonstrate environment-adaptive swarm geometry across different actuation and sensing settings.

	For mean-shift-based navigation, we consider how to coordinate environment-dependent morphology evolution with the swarm's ability to follow a moving reference. The objective is to accommodate both obstacle geometry and the redistribution needed to realize each morphology. We address this challenge through joint position--orientation--morphology planning, execution feedback, and moving-frame motion compensation.

	Figure~\ref{fig:framework_overview} summarizes the proposed target-field-based framework and its feedback from swarm execution to reference progression. The target field specifies an unlabeled spatial region without assigning robots to individual target cells. The planner searches a discretized position--orientation--morphology space, checking candidate fields and sampled transitions against workspace and obstacle constraints. The reference manager converts the discrete sequence into a continuous pose reference with discrete morphology switches. Swarm entry, coverage, and lag measurements regulate reference progression and switching. A moving-frame controller combines prescribed rigid motion with local mean-shift redistribution, neighbor interactions, and obstacle responses. This architecture connects environment-dependent target-field planning with assignment-free physical execution.

	The main contributions of this paper are summarized as follows.
	\begin{itemize}
		\item An environment-adaptive target-field formulation extends assignment-free shape assembly to obstacle-constrained navigation. A planner jointly selects field position, orientation, and morphology using geometric feasibility checks, without introducing robot-specific goals. A clearance-based sufficient condition relates edge sampling to continuous target-field feasibility.

		\item A feedback-regulated reference manager connects discrete field planning to swarm execution. Assignment-free entering, coverage, and lag metrics regulate reference progression, while pre- and post-switch guards coordinate morphology updates with local redistribution.

		\item A moving-frame mean-shift controller combines field translation and rotation with assignment-free redistribution. Kinematic analysis establishes nominal frame-motion cancellation, a sufficient condition for unsaturated execution, and a bound on local-motion deviations. Simulations with 10--100 robots and three-robot hardware validation demonstrate the integrated framework, while a fixed-ring ablation examines the role of morphology selection.
	\end{itemize}

	Here, ``morphology-preserving'' denotes the objective of maintaining coherent swarm occupation of an evolving unlabeled target field. Individual robots can redistribute as the reference moves and the morphology changes.

	The remainder of this paper is organized as follows. Section~\ref{sec:problem_formulation} states the navigation task and objectives, then defines the swarm model, target-field representation, feasibility conditions, and execution metrics. Section~\ref{sec:target_field_generation} presents the feasibility-aware target-field planner and execution-regulated reference manager.  Section~\ref{sec:mean_shift} develops the moving target-field frame and morphology-preserving mean-shift navigation law. Section~\ref{sec:experiments} reports the simulation and experimental evaluation, and Section~\ref{sec:conclusion} concludes the paper.

	\section{Problem Formulation}
	\label{sec:problem_formulation}

	\subsection{Navigation Task and Objectives}
	\label{subsec:navigation_task_objectives}
	Consider $N$ mobile robots navigating in a known workspace with static obstacles. Given their initial configuration, an initial collective reference, a destination, and a finite library of candidate morphologies, the task is to guide the swarm to the destination while adapting its shape to the available free space.

	We seek a target-field trajectory $\mathcal{T}_d(t)=(\mathbf{c}_d(t),\theta_d(t),k_d(t))$ and robot velocity commands $\{\mathbf{u}_i(t)\}_{i=1}^{N}$. The field center $\mathbf{c}_d$, orientation $\theta_d$, and morphology index $k_d$ describe a collective reference rather than robot-specific goals. The navigation objectives are:
	\begin{enumerate}
		\item \textit{Environment adaptation:} Select target-field positions, orientations, and morphologies that accommodate the workspace boundary and obstacle geometry along the route to the destination.
		\item \textit{Coordinated execution:} Coordinate reference progression and morphology changes with robot redistribution, while maintaining collective coherence and the desired robot clearances.
		\item \textit{Assignment-free navigation:} Form and follow the evolving target region without assigning individual robots to distinct target cells.
	\end{enumerate}
	The following subsections define the swarm and target-field representations, followed by the feasibility conditions and execution metrics used to specify these objectives.

	\subsection{Swarm and Target-Field Representation}
	\label{subsec:swarm_target_field_representation}

	Consider a team of $N$ homogeneous mobile robots operating in a bounded workspace $\Omega = [0,W] \times [0,H] \subset \mathbb{R}^{2}$. The environment contains $N_o$ static obstacles, whose union is denoted by $\Omega_o = \bigcup_{m=1}^{N_o}O_m$, where $O_m \subset \Omega$ is the $m$-th obstacle. Each robot is modeled as a disk of radius $r_b$.

	Let $\mathbf{p}_i = [x_i,y_i]^\top \in \mathbb{R}^2$ denote the position of robot $i$. The robot motion is described by the single-integrator kinematic model
	\begin{equation}
		\dot{\mathbf{p}}_i=\mathbf{u}_i, \qquad \|\mathbf{u}_i\|\le u_{\max}, \qquad i=1,\ldots,N ,
		\label{eq:robot_dynamics}        
	\end{equation}
	where $\mathbf{u}_i \in \mathbb{R}^2$ is the velocity command and $u_{\max}$ is the maximum admissible speed. The first-order model is used for swarm-level coordination and can be implemented on mobile platforms through low-level velocity tracking. The initial robot configuration is denoted by $\mathbf{P}(0)=[\mathbf{p}_1^\top(0),\ldots,\mathbf{p}_N^\top(0)]^\top$.

	The desired collective morphology is represented by a finite library of grid-based templates, denoted by $\mathcal{L} = \{\mathcal{M}^1, \mathcal{M}^2, \ldots, \mathcal{M}^K\}$. Each template $\mathcal{M}^k$ is defined in a local morphology frame as a two-dimensional binary or gray-level grid. In contrast to target-point-based formations, a template specifies a spatial target region rather than $N$ labeled target positions. For template $\mathcal{M}^k$, let the grid size be $N_r^k \times N_c^k$. A cell is indexed by $\nu = (r, c)$, where $r \in \{0, \ldots, N_r^k-1 \}$ and $c \in \{0, \ldots, N_c^k-1\}$. Let $s_\nu^k \in \{0, 1\}$ denote the binary value of cell $\nu$, where $s_\nu^k = 0$ indicates that the cell belongs to the target region, and $s_\nu^k = 1$ indicates that the cell lies outside the target region. The set of target-region cells can be denoted by $\mathcal{B}_k = \{\nu \mid s_\nu^k = 0\}$. The binary template is further converted into a gray-level guidance field to enlarge the attraction range of the target region. For each cell $\nu$ in template $\mathcal{M}^k$, the guidance value is denoted by $\Gamma_k(\nu) \in [0,1]$. The field is generated from the binary template through a distance transformation, so that cells closer to the target region are assigned lower guidance values. The resulting gray-level guidance field is used by the mean-shift controller to guide robots smoothly toward the desired morphology, whereas the binary set $\mathcal{B}_k$ is used by the planning layer for geometric feasibility checking.  

	To maintain comparable physical density across templates, each morphology is assigned a template-dependent cell length. Let $n_c^k = |\mathcal{B}_k|$ be the number of target-region cells and let $r_a$ denote the interaction radius between robots. The physical side length of each square cell is defined as 
	\begin{equation}
		l_c^k = \sqrt{\frac{\pi N}{4n_c^k}}\, r_a.
		\label{eq:cell_length}
	\end{equation}
	This scaling makes the target-region area proportional to the swarm size, maintaining a consistent nominal robot density across templates.

	For each target cell $\nu \in \mathcal{B}_k$, its local center coordinate in the morphology frame is defined as 
	\begin{equation}
		\mathbf{x}_{k,\nu}^L =
		\begin{bmatrix}
			\left(c+\frac{1}{2}-\frac{N_c^k}{2}\right)l_c^k
			\\
			\left(r+\frac{1}{2}-\frac{N_r^k}{2}\right)l_c^k
		\end{bmatrix}.
	\end{equation}
	The superscript $L$ indicates that this coordinate is expressed in the local frame attached to the morphology center. Thus, each template is defined independently of its global position and orientation. During navigation, the target-field state will translate and rotate these local cell centers into the world frame, allowing the same morphology template to be reused at different locations, headings, and switching stages. For later use, we define the local target-cell center set of morphology $k$ as $\mathcal{X}_k^L = \{\mathbf{x}_{k,\nu}^L \mid \nu\in \mathcal{B}_k\}$.

	The morphology library introduced above defines a set of local spatial patterns. To use these patterns for navigation, each morphology must be embedded into the workspace with a specified position and orientation. We therefore introduce the target field as a virtual, unlabeled spatial reference for the entire swarm. A target field is parameterized by 
	\begin{equation}
		\mathcal{T} = (\mathbf{c}, \theta, k),
		\label{eq:target_field}
	\end{equation}
	where $\mathbf{c} = [c_x, c_y]^\top \in \mathbb{R}^2$ is the field center, $\theta \in (-\pi, \pi]$ is the field orientation, and $k \in \{1,\ldots, K\}$ selects the morphology template. The elements of $\mathcal{B}_k$ are referred to as target cells in the remainder of the paper. These cells describe the desired occupied region of the swarm, but they are not assigned to individual robots. The initial field is $\mathcal{T}_0=(\mathbf{c}_0,\theta_0,k_0)$, and the goal center is $\mathbf{c}_{\rm g}$. 

	The planar rotation matrix can be denoted by
	\begin{equation}
		R(\theta)
		=
		\begin{bmatrix}
			\cos\theta & -\sin\theta\\
			\sin\theta &  \cos\theta
		\end{bmatrix}.
		\label{eq:rotation_matrix}
	\end{equation}
	For a target cell $\nu \in \mathcal{B}_k$, the corresponding world-frame cell center is obtained by rigidly transforming its local coordinate:
	\begin{equation}
		\mathbf{x}_{k,\nu}^W(\mathcal{T}) = 
		\mathbf{c} + R(\theta) \mathbf{x}_{k,\nu}^L.
		\label{eq:world_cell_coordinate}
	\end{equation}
	Accordingly, the world-frame target-cell center set induced by $\mathcal{T}$ is $\mathcal{X}_k^W(\mathcal{T}) = \{\mathbf{c} + R(\theta) \mathbf{x} \mid \mathbf{x}\in \mathcal{X}_k^L\}$. For robot-level computation, robot positions are expressed in the current target-field frame as 
	\begin{equation}
		\mathbf{p}_i^{L}
		=
		R(\theta)^\top(\mathbf{p}_i-\mathbf{c}).
		\label{eq:robot_local_coordinate}
	\end{equation}
	This coordinate transformation allows the same local morphology template and gray-level field to be used when the target field translates, rotates, and switches its morphology. Since the target field represents an unlabeled spatial distribution, no one-to-one correspondence is introduced between robots and target cells.

\begin{figure}[t]
		\centering
		\includegraphics[width=0.98\columnwidth]{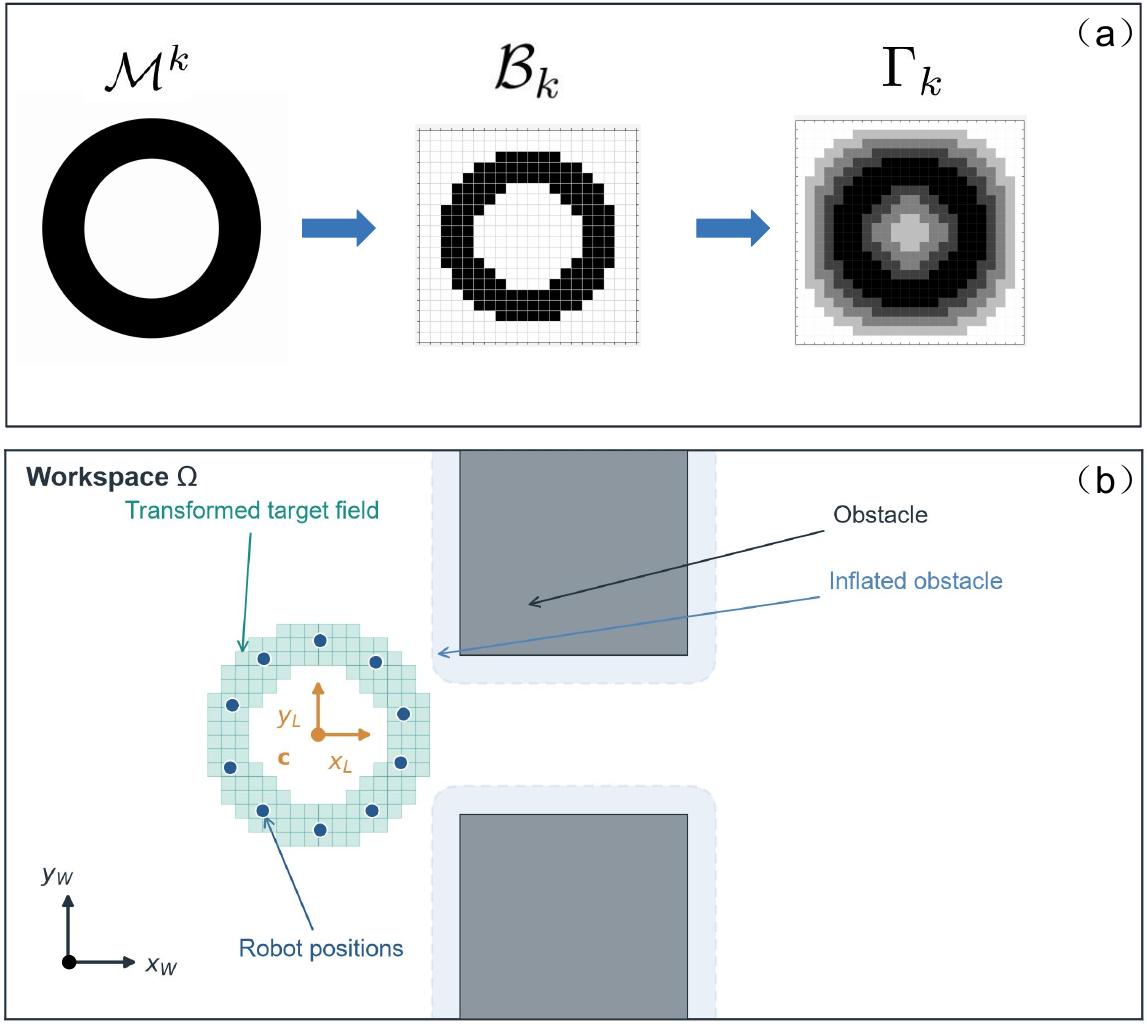}
		\caption{Morphology representation and its embedding in the workspace. (a) The ring-shaped template $\mathcal{M}^{k}$ is represented by the target-cell set $\mathcal{B}_{k}$ and converted into the gray-level guidance field $\Gamma_{k}$, which is zero on target cells and increases with grid distance from the target region. (b) The target field $\mathcal{T}=(\mathbf{c},\theta,k)$ embeds the target cells in workspace $\Omega$ through translation and rotation. Blue dashed contours indicate the planner-inflated obstacles $\bar{O}_m=O_m\oplus\mathcal{B}(0,d_p)$. Robot positions are shown within the unlabeled target field, without persistent robot-to-cell assignments.}
		\label{fig:representations_and_frames}
	\end{figure}

	Fig.~\ref{fig:representations_and_frames} illustrates the grid-based morphology, its gray-level guidance field, and the translation--rotation mapping from local target cells to the workspace.

	\textit{Remark 1:} The proposed formulation differs from target-point-based formation control in that each morphology template specifies a spatial region rather than a set of labeled goals. Consequently, no Hungarian assignment, auction assignment, or local task swapping is required during morphology generation or switching.

	\subsection{Feasibility Conditions and Execution Metrics}
	\label{subsec:feasibility_execution_metrics}

	For geometric reasoning in the planning layer, obstacles are inflated by a prescribed planning margin $d_p > 0$. The inflated obstacle set is $\bar{O}_m = O_m \oplus \mathcal{B}(0, d_p)$, where $\oplus$ denotes the Minkowski sum and $\mathcal{B}(0,d_p)$ is the closed disk of radius $d_p$ centered at the origin. The inflated obstacle region can be denoted by $\bar{\Omega}_o = \bigcup_{m=1}^{N_o}\bar{O}_m$. The free space available for planning is defined as $\bar{\Omega}_f = \Omega \setminus \bar{\Omega}_o$.

	The target field also provides the geometric object used by the planning layer. A target field is regarded as feasible if its target-cell region lies inside the workspace and does not intersect the inflated obstacles. Because feasibility is checked using cell centers, the cell half-diagonal $\delta_k = \sqrt{2}l_c^k/2$ accounts for the finite extent of each cell. The effective planning clearance of template $k$ is then defined as $d_k = d_p +\delta_k$. For every target-cell center $\mathbf{x} = [x, y]^\top \in \mathcal{X}_k^W(\mathcal{T})$, boundary feasibility requires 
	\begin{equation}
		d_k \le x \le W-d_k, \qquad d_k \le y \le H-d_k .
		\label{eq:boundary_feasibility}
	\end{equation}
	Obstacle feasibility requires that the transformed target-cell set does not intersect any obstacle inflated by the template-dependent clearance:
	\begin{equation}
		\mathcal{X}_{k}^{W}(\mathcal{T}) \cap \left(O_m\oplus\mathcal{B}(0,d_k)\right) = \emptyset,
		\qquad
		m=1,\ldots,N_o .
		\label{eq:obstacle_feasibility}
	\end{equation}
	The minimum distance from the target-cell centers to the obstacles is defined as 
	\begin{equation}
		d_{\min}(\mathcal{T}) = \min_{\mathbf{x}\in\mathcal{X}_{k}^{W}(\mathcal{T})} \min_{m=1,\ldots,N_o} \operatorname{dist}(\mathbf{x},O_m),
		\label{eq:target_field_clearance}
	\end{equation}
	where $\operatorname{dist}(\mathbf{x},O_m)=\inf_{\mathbf{y}\in O_m}\|\mathbf{x}-\mathbf{y}\|$. For the closed inflated obstacles, $d_{\min}(\mathcal{T}) > d_k$ gives a sufficient condition for obstacle feasibility. The feasible target-field set is therefore defined as 
	\begin{equation}
		\mathbb{T}_{\rm free} = \left\{ \mathcal{T} \mid \mathcal{T}
		\ \text{satisfies}
		\ \eqref{eq:boundary_feasibility}
		\ \text{and}
		\ \eqref{eq:obstacle_feasibility} \right\}.
		\label{eq:feasible_target_field_set}
	\end{equation}
	The feasibility notion is imposed on the virtual target field rather than on the instantaneous multi-robot configuration. Thus, a feasible target field indicates that the desired collective morphology can be geometrically placed in the environment, while the actual robots may still deviate from the field during entering, exploration, local interaction, or morphology transition.

	To regulate the motion of the target field according to the physical execution state of the swarm, we introduce three assignment-free metrics: entering ratio, coverage ratio, and lag error. These metrics evaluate the relationship between the robots and the current target field, but they do not assign any robot to any target cell. Their dependence on the current robot configuration is suppressed in the notation.

	A robot is considered to have entered the target field if its position is within a tolerance $d_{\rm in} > 0$ of the local target-cell center set. The entering indicator of robot $i$ is
	\begin{equation}
		\chi_i(\mathcal{T}) = \mathbb{I} \left[ \min_{\mathbf{x}\in\mathcal{X}_{k}^{L}} \|\mathbf{p}_i^{L}-\mathbf{x}\| \le d_{\rm in} \right],
		\label{eq:entering_indicator}
	\end{equation}
	where $\mathbb{I}[\cdot]$ denotes the indicator function, which returns 1 if the statement inside the brackets is true and 0 otherwise. Therefore, $\chi_i(\mathcal{T}) = 1$ means that robot $i$ has entered the target field, while $\chi_i(\mathcal{T}) = 0$ means that it has not. The entering ratio is defined as 
	\begin{equation}
		M_{\rm ent}(\mathcal{T}) = \frac{1}{N}\sum_{i=1}^{N}\chi_i(\mathcal{T}).
		\label{eq:entering_ratio}
	\end{equation}
	This metric measures the fraction of robots that have reached the target-field region.

	Coverage evaluates whether the target cells are sufficiently occupied by the swarm. A target cell is considered covered if at least one robot lies within an occupation radius $d_{\rm occ} > 0$ of its local cell center $\mathbf{x}_{k,\nu}^L$ in the target-field frame. The coverage ratio is defined as 
	\begin{equation}
		M_{\rm cov}(\mathcal{T}) = 
		\frac{1}{|\mathcal{B}_k|} \sum_{\nu\in\mathcal{B}_k} \mathbb{I} \left[ d_\nu^{\rm cov}(\mathcal{T}) \le d_{\rm occ} \right],
		\label{eq:coverage_ratio}
	\end{equation}
	where $d_\nu^{\rm cov}(\mathcal{T}) = \min_{i=1,\ldots,N} \|\mathbf{p}_i^L-\mathbf{x}_{k,\nu}^{L}\|$. Unlike target assignment, this definition only checks whether a target cell is covered by at least one robot, regardless of its identity.

	Finally, the lag error measures the translational mismatch between the swarm and the target field:
	\begin{equation}
		e_{\rm lag}(\mathcal{T}) = \|\mathbf{p}_c-\mathbf{c}\|,
		\label{eq:lag_error}
	\end{equation}
	where $\mathbf{p}_c = \frac{1}{N}\sum_{i=1}^{N}\mathbf{p}_i$ denotes the swarm centroid. The entering ratio, coverage ratio, and lag error are used by the reference manager to determine the motion mode of the target field, including nominal motion, speed reduction, waiting, and delayed morphology switching.

	Although the target field is planned to avoid obstacles at the morphology level, individual robots may temporarily move outside the target-cell region during execution. Therefore, robot-level clearance remains an execution requirement. Let $d_{\rm r} > 2r_b$ be the desired minimum center-to-center distance between robots. Let $d_{\rm o} > r_b$ be the desired minimum distance from a robot center to an obstacle. The desired clearance conditions are
	\begin{equation}
		\|\mathbf{p}_i-\mathbf{p}_j\|\ge d_{\rm r}, \qquad \forall i\ne j,
		\label{eq:robot_robot_clearance}
	\end{equation}
	and
	\begin{equation}
		\operatorname{dist}(\mathbf{p}_i,O_m)\ge d_{\rm o},
		\label{eq:robot_obstacle_clearance}
	\end{equation}
	for $i=1,\ldots,N$ and $m=1,\ldots,N_o$. In the proposed framework, these clearance requirements are handled at the execution layer through local inter-robot interaction and obstacle-avoidance terms in the mean-shift navigation law.

	Using these definitions, the desired navigation objectives introduced in Section~\ref{subsec:navigation_task_objectives} are expressed as 
	\begin{equation}
		\begin{aligned}
			&\mathcal{T}_d(t) \in \mathbb{T}_{\rm free}, \qquad \forall t \in [0, T_f],\\
			&\| \mathbf{c}_d(T_f) - \mathbf{c}_{\rm g} \| \le \varepsilon_{\rm g},\\
			&M_{\rm cov}(\mathcal{T}_d(t))\ge \eta_{\rm cov}, \quad
			M_{\rm ent}(\mathcal{T}_d(t))\ge \eta_{\rm ent}.
		\end{aligned}
	\end{equation}
	The conditions in the last line are desired performance targets during non-switching motion, excluding initial assembly and morphology-transition transients. Robot-level execution additionally seeks to satisfy the clearance requirements in \eqref{eq:robot_robot_clearance} and \eqref{eq:robot_obstacle_clearance} through local interaction-based navigation. Here, $T_f$ is the terminal time, $\varepsilon_{\rm g}$ is the goal tolerance, and $\eta_{\rm cov}, \eta_{\rm ent} \in (0,1]$ are desired coverage and entering thresholds.

	\textit{Remark 2:} The feasibility of the virtual target field does not imply that every robot remains inside the field at every instant. The role of the target field is to provide an adaptive collective reference, while robot-level motion is generated by assignment-free mean-shift navigation.

	\section{Adaptive Target-Field Generation}
	\label{sec:target_field_generation}
\begin{figure*}[t]
		\centering
		\includegraphics[ width=\textwidth,
		trim=2mm 8mm 4mm 8mm,
		clip]{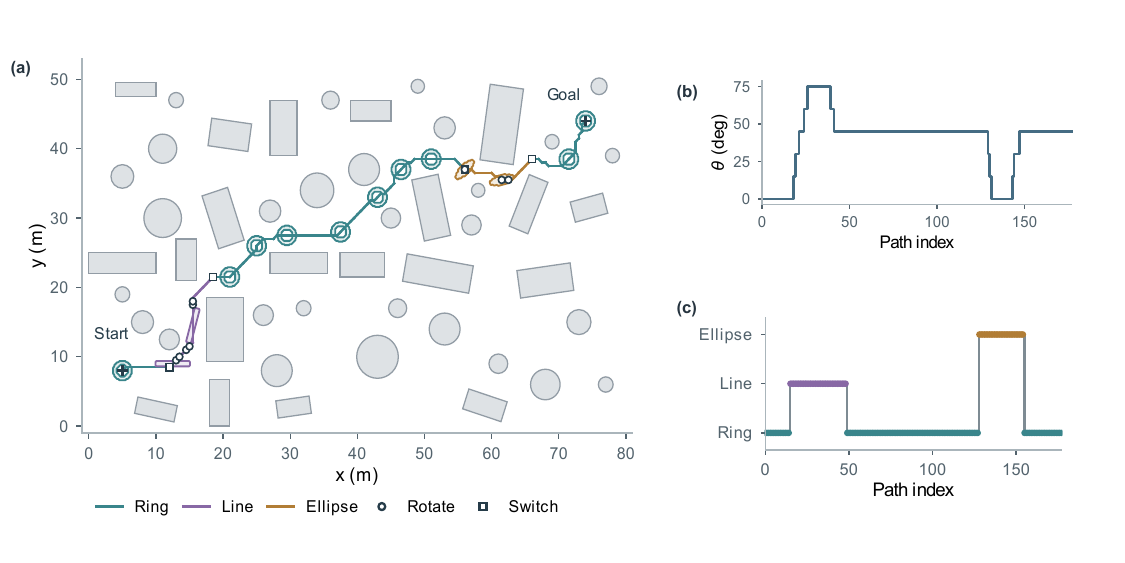}
		\caption{Planning result in the obstacle environment. (a) Planned target-field route; colored outlines indicate representative morphologies, circles mark rotation actions, and squares mark morphology switches. (b) Planned heading along the discrete path. (c) Selected morphology sequence, in which the target field changes from a ring to a line, returns to a ring, adopts an ellipse, and recovers the ring at the goal.}
		\label{fig:planning_result}
	\end{figure*}

	The method has two linked tasks: choose a geometrically admissible target-field route and coordinate its execution with the swarm. The planner searches over position, orientation, and morphology, while the reference manager controls temporal progress and switching. Fig.~\ref{fig:planning_result} illustrates the resulting route, heading changes, and morphology sequence.

	\subsection{Target-Field State Graph}
	\label{subsec:target_field_state_graph}
	Each graph node describes a collective reference, so the planning variables remain independent of individual robot identities. With spatial resolution $\Delta_p>0$ and $N_\theta$ headings, the discrete state and its associated target field are
	\begin{equation}
		\begin{aligned}
			q&=(i_x,i_y,i_\theta,k),\\
			\mathcal{T}(q)&=(\mathbf{c}_q,\theta_q,k_q).
		\end{aligned}
		\label{eq:target_field_graph_state}
	\end{equation}
	Here, $i_x\in\mathcal{I}_x$ and $i_y\in\mathcal{I}_y$ index the workspace grid, $i_\theta\in\{0,\ldots,N_\theta-1\}$ indexes orientation, and $k\in\{1,\ldots,K\}$. The mapping uses $\mathbf{c}_q=[i_x\Delta_p,i_y\Delta_p]^\top$, $\theta_q=\operatorname{wrap}(-\pi+i_\theta\Delta_\theta)$, $\Delta_\theta=2\pi/N_\theta$, and $k_q=k$. The operator $\operatorname{wrap}$ maps angles to $(-\pi,\pi]$. All candidate orientations are retained here; rotationally symmetric templates may alternatively use a reduced interval.

	Let $\mathcal{Q}$ contain all discrete states. A state is retained only when its target field satisfies the geometric conditions in Section~\ref{subsec:feasibility_execution_metrics}:
	\begin{equation}
		\mathcal{V} = \left\{q \in \mathcal{Q} \mid \mathcal{T}(q) \in \mathbb{T}_{\rm free} \right\}.
		\label{eq:graph_vertex_set}
	\end{equation}
	This vertex test determines where each morphology can be placed. Three elementary actions then connect candidate states: translation moves the field, rotation aligns it with a passage, and switching changes its template. Their successors are collected as
	\begin{equation}
		q'=\begin{cases}
			(i_x+\delta_x,i_y+\delta_y,i_\theta,k), &\text{translation},\\
			(i_x,i_y,\operatorname{wrap}_{N_\theta}(i_\theta+\sigma),k), &\text{rotation},\\
			(i_x,i_y,i_\theta,k'), &\text{switching}.
		\end{cases}
		\label{eq:target_field_actions}
	\end{equation}
	Translations are eight-connected: $\delta_x,\delta_y\in\{-1,0,1\}$ and $(\delta_x,\delta_y)\ne(0,0)$. Rotations use $\sigma\in\{-1,1\}$, with $\operatorname{wrap}_{N_\theta}$ wrapping the index to $\{0,\ldots,N_\theta-1\}$. Switching uses $k'\in\mathcal{K}_{\rm sw}(k)\subseteq\{1,\ldots,K\}\setminus\{k\}$. This set permits all other templates unless the library restricts particular transitions. Translation and rotation preserve $k$; switching preserves the field center and orientation.

	Denote the three successor sets by $\mathcal{N}_{\rm tr}(q)$, $\mathcal{N}_{\rm rot}(q)$, and $\mathcal{N}_{\rm sw}(q)$. Their union is $\mathcal{N}_{\rm cand}(q)$. Candidate edges join feasible endpoint states:
	\begin{equation}
		\mathcal{E}_{\rm cand} = \left\{ (q,q') \mid q\in\mathcal{V},\ q'\in\mathcal{N}_{\rm cand}(q)\cap\mathcal{V} \right\}.
		\label{eq:candidate_edge_set}
	\end{equation}
	The initial state is $q_{\rm s}=\Pi_{\mathcal{Q}}(\mathcal{T}_0)$, the nearest discrete representation of the initial field. It is accepted only if $q_{\rm s}\in\mathcal{V}$; otherwise, the initial field is adjusted to the nearest feasible state. With goal tolerance $\varepsilon_{\rm g}>0$, define $\mathcal{V}_{\rm g}=\{q\in\mathcal{V}\mid\|\mathbf{c}_q-\mathbf{c}_{\rm g}\|\le\varepsilon_{\rm g}\}$. Transition validation, described next, reduces $\mathcal{E}_{\rm cand}$ to the search edge set $\mathcal{E}$.

	\subsection{Feasibility-Aware Transition Search}
	\label{subsec:feasibility_aware_transition_search}
	The central planning problem is to select a route and morphology sequence together. On the validated graph $\mathcal{G}=(\mathcal{V},\mathcal{E})$, let $Q=(q_0,\ldots,q_T)$ be an ordered path with $T$ edges. The search solves
	\begin{equation}
		\begin{aligned}
			Q^\star&=\mathop{\arg\min}_{Q}\sum_{l=0}^{T-1}C(q_l,q_{l+1})\\
			\text{s.t.}\quad & q_0=q_{\rm s},\quad q_T\in\mathcal{V}_{\rm g},\\
			&(q_l,q_{l+1})\in\mathcal{E},\quad l=0,\ldots,T-1.
		\end{aligned}
		\label{eq:target_field_shortest_path}
	\end{equation}
	Feasibility determines the admissible transitions; the cost ranks the remaining alternatives. The following two parts specify these roles separately.

	\textit{1) Transition validation:} Feasible endpoints alone do not establish that an intervening field avoids obstacles. For a candidate edge, interpolation at $\lambda\in[0,1]$ gives
	\begin{equation}
		\begin{aligned}
			\mathbf{c}_\lambda&=(1-\lambda)\mathbf{c}_q+\lambda\mathbf{c}_{q'},\\
			\theta_\lambda&=\operatorname{wrap}\!\left(\theta_q+\lambda\operatorname{wrap}(\theta_{q'}-\theta_q)\right),\\
			\mathcal{T}_\lambda(q,q')&=(\mathbf{c}_\lambda,\theta_\lambda,k_\lambda).
		\end{aligned}
		\label{eq:interpolated_target_field}
	\end{equation}
	For translation and rotation, $k_\lambda=k_q=k_{q'}$. With an integer $L_e\ge1$ and samples $\lambda_l=l/L_e$, non-switching edges must satisfy
	\begin{equation}
		\mathcal{T}_{\lambda_l}(q,q')\in\mathbb{T}_{\rm free}, \qquad
		l=0,\ldots,L_e .
		\label{eq:sampled_edge_feasibility}
	\end{equation}
	For switching, the center and orientation are fixed, and both templates must fit at that pose:
	\begin{equation}
		(\mathbf{c}_q,\theta_q,k_q)\in\mathbb{T}_{\rm free},
		\qquad
		(\mathbf{c}_q,\theta_q,k_{q'})\in\mathbb{T}_{\rm free}.
		\label{eq:switching_feasibility}
	\end{equation}
	The search therefore retains edges passing their action-specific check:
	\begin{equation}
		\mathcal{E}=\left\{(q,q')\in\mathcal{E}_{\rm cand} \mid
		\text{validation passed} \right\}.
		\label{eq:validated_edge_set}
	\end{equation}
	These checks reject sampled intermediate collisions and incompatible template updates. They are particularly relevant to elongated morphologies, whose rotation can sweep a large region even when the center remains fixed.

	The spacing of validation samples can also be related to continuous feasibility. For a non-switching edge with fixed template $k$, define $\Delta\mathbf{c}_e=\mathbf{c}_{q'}-\mathbf{c}_q$, $\Delta\theta_e=\operatorname{wrap}(\theta_{q'}-\theta_q)$, and $\rho_k=\max_{\nu\in\mathcal{B}_k}\|\mathbf{x}_{k,\nu}^L\|$. Each interpolated cell center differs from its counterpart at the nearest validation parameter by at most
	\begin{equation}
		\varepsilon_{\rm edge}=
		\frac{\|\Delta\mathbf{c}_e\|+\rho_k|\Delta\theta_e|}{2L_e}.
		\label{eq:edge_sampling_displacement_bound}
	\end{equation}

	\begin{proposition}[Continuous feasibility of a sampled edge]
		\label{prop:continuous_edge_feasibility}
		Consider the non-switching interpolation in \eqref{eq:interpolated_target_field} with uniform samples $\lambda_j=j/L_e$, $j=0,\ldots,L_e$. Write $[x_{\nu,j},y_{\nu,j}]^\top=\mathbf{x}_{k,\nu}^W(\mathcal{T}_{\lambda_j})$. If there exists a margin $\mu_e>\varepsilon_{\rm edge}$ such that every sample satisfies
		\begin{equation}
			\begin{aligned}
				d_{\min}(\mathcal{T}_{\lambda_j})&\ge d_k+\mu_e,\\
				d_k+\mu_e\le x_{\nu,j}&\le W-d_k-\mu_e,\\
				d_k+\mu_e\le y_{\nu,j}&\le H-d_k-\mu_e,
			\end{aligned}
			\label{eq:continuous_edge_margin_condition}
		\end{equation}
		for all $\nu\in\mathcal{B}_k$ and $j=0,\ldots,L_e$, then $\mathcal{T}_\lambda(q,q')\in\mathbb{T}_{\rm free}$ for every $\lambda\in[0,1]$.
	\end{proposition}

	\begin{IEEEproof}
		For any $\lambda$, choose its nearest sample $\lambda_j$, so $|\lambda-\lambda_j|\le1/(2L_e)$. Since $\|R(a)-R(b)\|_2\le|a-b|$, the interpolated cell centers satisfy
		\begin{equation*}
			\begin{aligned}
				&\|\mathbf{x}_{k,\nu}^W(\mathcal{T}_\lambda)
				-\mathbf{x}_{k,\nu}^W(\mathcal{T}_{\lambda_j})\|\\
				&\quad\le |\lambda-\lambda_j|
				\bigl(\|\Delta\mathbf{c}_e\|+\rho_k|\Delta\theta_e|\bigr)
				\le\varepsilon_{\rm edge}.
			\end{aligned}
		\end{equation*}
		The rotation bound applies to continuous angular representatives, so wrapping does not affect this estimate. Distance to an obstacle changes by at most the cell-center displacement. Hence every intermediate center has obstacle distance at least $d_k+\mu_e-\varepsilon_{\rm edge}>d_k$. Each coordinate changes by at most the same bound, preserving \eqref{eq:boundary_feasibility}. Both feasibility conditions therefore hold throughout the edge.
	\end{IEEEproof}

	Proposition~\ref{prop:continuous_edge_feasibility} gives a sufficient relation between sample spacing and additional clearance. The basic test in \eqref{eq:sampled_edge_feasibility} does not enforce this extra margin; continuous feasibility is certified only for edges satisfying \eqref{eq:continuous_edge_margin_condition}. The result concerns the virtual field on non-switching edges, while template switching and robot-level motion require their respective execution checks.

	\textit{2) Transition cost and search:} The cost balances motion, switching, template preference, and obstacle clearance:
	\begin{equation}
		\begin{aligned}
			C(q,q') = & C_{\rm tr}(q,q') + C_{\rm rot}(q,q') + C_{\rm sw}(q,q') \\
			& + C_{\rm temp}(q') + C_{\rm clr}(q').
		\end{aligned}
		\label{eq:transition_cost_total}
	\end{equation}
	Rather than introducing a separate planning problem for each preference, all five terms contribute to the same edge cost:
	\begin{equation}
		\begin{aligned}
			C_{\rm tr}(q,q')&=w_{\rm tr}\|\mathbf{c}_{q'}-\mathbf{c}_q\|,\\
			C_{\rm rot}(q,q')&=w_{\rm rot}|\operatorname{wrap}(\theta_{q'}-\theta_q)|,\\
			C_{\rm sw}(q,q')&=w_{\rm sw}\mathbb{I}[k_{q'}\ne k_q],\\
			C_{\rm temp}(q')&=w_{\rm temp}\eta_{k_{q'}},\\
			C_{\rm clr}(q')&=w_{\rm clr}\left[\max\{0,d_{\rm pref}-\hat d(q')\}\right]^2.
		\end{aligned}
		\label{eq:transition_cost_components}
	\end{equation}
	Here, $w_{\rm tr}>0$ and the other weights are nonnegative. The first three terms penalize travel, rotation, and template changes. The nonnegative prior $\eta_{k_{q'}}$ favors selected templates. In the clearance term, $\hat d(q')$ is the minimum obstacle and workspace-boundary clearance of the finite-cell target-field footprint, and $d_{\rm pref}>0$ is the preferred clearance. The squared penalty is active only when $\hat d(q')<d_{\rm pref}$; hard feasibility is checked separately.

	Assigning lower prior costs to nominal compact templates encourages their use in open space and their recovery after constrained passages. A narrow passage can instead favor an elongated template when the compact one is infeasible or a detour costs more. Thus, the morphology sequence is selected by the same route search, not prescribed as a switching schedule.

	A* solves \eqref{eq:target_field_shortest_path}. Its evaluation is $f(q)=g(q)+h(q)$, where $g(q)$ is the accumulated cost from $q_{\rm s}$. The heuristic is the weighted distance to the goal-center region:
	\begin{equation}
		h(q)=w_{\rm tr} \max\left\{ 0,\, \|\mathbf{c}_q-\mathbf{c}_{\rm g}\| -\varepsilon_{\rm g} \right\}.
		\label{eq:target_field_heuristic}
	\end{equation}
	Algorithm~\ref{alg:adaptive_target_field_search} generates candidate successors and validates vertices and edges on demand, caching geometric checks during search rather than constructing the full graph in advance. The returned path specifies collective field states without assigning robots to target cells.

	\begin{algorithm}[t]
		\caption{Feasibility-Aware Target-Field Search}
		\label{alg:adaptive_target_field_search}
		\begin{algorithmic}[1]
			\REQUIRE $\mathcal{L}$, $\{O_m\}_{m=1}^{N_o}$, $q_{\rm s}$, $\mathcal{V}_{\rm g}$
			\ENSURE Discrete target-field path $Q^\star$
			\STATE Initialize empty vertex- and edge-validation caches
			\STATE Initialize the open set with $q_{\rm s}$
			\STATE Set $g(q_{\rm s})=0$ and ${\rm parent}(q_{\rm s})=\emptyset$
			\STATE Use $g(q)=+\infty$ for each unseen state
			\WHILE{the open set is not empty}
			\STATE Select and remove $q$ with minimum $g(q)+h(q)$
			\IF{$q\in\mathcal{V}_{\rm g}$}
			\STATE Reconstruct $Q^\star$ from parent pointers
			\RETURN $Q^\star$
			\ENDIF
			\FOR{each $q'\in\mathcal{N}_{\rm cand}(q)$}
			\STATE Query or cache the checks in \eqref{eq:graph_vertex_set} and \eqref{eq:validated_edge_set}
			\IF{$q'$ and $(q,q')$ pass their checks}
			\STATE $g_{\rm new}\leftarrow g(q)+C(q,q')$
			\IF{$g_{\rm new}<g(q')$}
			\STATE $g(q')\leftarrow g_{\rm new}$
			\STATE ${\rm parent}(q')\leftarrow q$
			\STATE Insert or update $q'$ in the open set
			\ENDIF
			\ENDIF
			\ENDFOR
			\ENDWHILE
			\RETURN failure
		\end{algorithmic}
	\end{algorithm}

	\textit{Remark 3:} Algorithm~\ref{alg:adaptive_target_field_search} returns feasible graph vertices and transitions satisfying their action-specific checks. Non-switching edges satisfy \eqref{eq:sampled_edge_feasibility}, while switching edges satisfy the two-template condition in \eqref{eq:switching_feasibility}. Proposition~\ref{prop:continuous_edge_feasibility} additionally certifies a non-switching edge when the margin condition holds. Without that condition, feasibility is established only at the checked samples. Neither the edge certificate nor the switching check guarantees robot-level collision avoidance.

	\subsection{Execution-Regulated Field Reference}
	\label{subsec:execution_regulated_field_reference}
	The path $Q^\star$ determines where the field should move and which morphologies it should visit. Execution regulation determines how quickly it advances and when a template change is released. Fixed timing can separate the virtual reference from robots delayed by interactions, obstacles, or redistribution. The manager therefore uses the entering, coverage, and lag metrics from Section~\ref{subsec:feasibility_execution_metrics} to coordinate two modes: non-switching motion and switching with redistribution.

	\textit{1) Advance, slow, or hold:} For an active non-switching edge $(q_l,q_{l+1})$, let $s(t)\in[0,1]$ be progress, with $s(t_l)=0$. A nominal edge duration $\tau_l>0$ is scaled by the execution-dependent factor $\alpha(t)$:
	\begin{equation}
		\begin{aligned}
			\dot{s}(t)&=\frac{\alpha(t)}{\tau_l},\\
			\alpha(t)&=\begin{cases}
				1,&\phi(t)\ge\phi_{\rm nom},\\
				\alpha_{\rm slow},&\phi_{\rm hold}<\phi(t)<\phi_{\rm nom},\\
				0,&\phi(t)\le\phi_{\rm hold}.
			\end{cases}
		\end{aligned}
		\label{eq:reference_speed_modulation}
	\end{equation}
	Here, $0<\phi_{\rm hold}<\phi_{\rm nom}\le1$ and $0<\alpha_{\rm slow}<1$. Progress is integrated from $t_l$ and capped at one. Thus, the path ordering is retained while execution can proceed nominally, slow down, or wait.

	The normalized score $\phi(t)$ measures the least-satisfied execution requirement:
	\begin{equation}
		\begin{aligned}
			\phi(t) = \min\Bigg\{ &1,\, \frac{M_{\rm ent}\bigl(\mathcal{T}_d(t)\bigr)}{\eta_{\rm ent}},\, \frac{M_{\rm cov}\bigl(\mathcal{T}_d(t)\bigr)}{\eta_{\rm cov}},
			\\ &\frac{e_{\max}}{e_{\rm lag}\bigl(\mathcal{T}_d(t)\bigr)+\varepsilon_e}\Bigg\}.
		\end{aligned}
		\label{eq:execution_score}
	\end{equation}
	The thresholds satisfy $0<\eta_{\rm ent},\eta_{\rm cov}\le1$ and $e_{\max}>0$; $\varepsilon_e>0$ is a distance regularizer. A score near one indicates adequate field entry and coverage with acceptable centroid lag. A lower score slows or holds the reference through \eqref{eq:reference_speed_modulation}.

	Progress parameterizes the pose and leaves the active template unchanged:
	\begin{equation}
		\begin{aligned}
			\mathbf{c}_d(t)&=(1-s(t))\mathbf{c}_{q_l}+s(t)\mathbf{c}_{q_{l+1}},\\
			\theta_d(t)&=\operatorname{wrap}\!\left(\theta_{q_l}+s(t)\Delta\theta_l\right),\\
			k_d(t)&=k_{q_l}=k_{q_{l+1}}.
		\end{aligned}
		\label{eq:continuous_center_reference}
	\end{equation}
	Here, $\Delta\mathbf{c}_l=\mathbf{c}_{q_{l+1}}-\mathbf{c}_{q_l}$ and $\Delta\theta_l=\operatorname{wrap}(\theta_{q_{l+1}}-\theta_{q_l})$. Let $\bar{\theta}_d$ be a continuous angular representative with $\theta_d=\operatorname{wrap}(\bar{\theta}_d)$. Differentiation on non-switching intervals gives the velocities supplied to the robot controller:
	\begin{equation}
		\dot{\mathbf{c}}_d(t) = \frac{\alpha(t)}{\tau_l} \Delta\mathbf{c}_l, \qquad
		\dot{\bar{\theta}}_d(t) = \frac{\alpha(t)}{\tau_l} \Delta\theta_l.
		\label{eq:target_field_velocity}
	\end{equation}
	The nominal duration enforces the prescribed field-speed bounds $v_{\rm f}>0$ and $\omega_{\rm f}>0$:
	\begin{equation}
		\tau_l = \max \left\{ \tau_{\min}, \frac{\|\Delta\mathbf{c}_l\|}{v_{\rm f}}, \frac{|\Delta\theta_l|}{\omega_{\rm f}} \right\},
		\label{eq:non_switching_edge_duration}
	\end{equation}
	The minimum duration $\tau_{\min}>0$ avoids excessively short segments. Translation and rotation are separate graph actions, although the same duration rule also accommodates simultaneous motion. The physical pose is continuous and piecewise differentiable, including when $\alpha$ changes mode.

	\textit{2) Switch and redistribute:} A switching edge has the same center and orientation at both endpoints but different template indices. The manager holds this pose and first checks
	\begin{equation}
		\begin{aligned}
			M_{\rm ent}(\mathcal{T}_d^-) &\ge \eta_{\rm ent}^{\rm pre}, \\
			M_{\rm cov}(\mathcal{T}_d^-) &\ge \eta_{\rm cov}^{\rm pre}, \\
			e_{\rm lag}(\mathcal{T}_d^-) &\le e_{\rm pre},
		\end{aligned}
		\label{eq:pre_switch_guard}
	\end{equation}
	Here, $\mathcal{T}_d^-$ is the pre-switch field, $\eta_{\rm ent}^{\rm pre},\eta_{\rm cov}^{\rm pre}\in(0,1]$, and $e_{\rm pre}>0$. If the guard fails, the field remains stationary with $\alpha=0$. Once it passes at $t_{\rm sw}$, the manager sets $k_d(t_{\rm sw}^+)=k_{q_{l+1}}$. The pose remains unchanged: $\mathbf{c}_d(t_{\rm sw}^+)=\mathbf{c}_d(t_{\rm sw}^-)$ and $\theta_d(t_{\rm sw}^+)=\theta_d(t_{\rm sw}^-)$. Superscripts $-$ and $+$ indicate values immediately before and after the update.

	Robots then redistribute relative to the new field $\mathcal{T}_d^+$. With a minimum dwell time $\tau_{\rm sw}\ge0$, progression resumes only when
	\begin{equation}
		\begin{aligned}
			t-t_{\rm sw} &\ge \tau_{\rm sw}, \\
			M_{\rm ent}(\mathcal{T}_d^+) &\ge \eta_{\rm ent}^{\rm post}, \\
			M_{\rm cov}(\mathcal{T}_d^+) &\ge \eta_{\rm cov}^{\rm post}, \\
			e_{\rm lag}(\mathcal{T}_d^+) &\le e_{\rm post}.
		\end{aligned}
		\label{eq:post_switch_release_guard}
	\end{equation}
	The post-switch thresholds satisfy $\eta_{\rm ent}^{\rm post},\eta_{\rm cov}^{\rm post}\in(0,1]$ and $e_{\rm post}>0$. The pre-switch guard checks readiness to change morphology; the post-switch guard allows redistribution before departure.

	\begin{algorithm}[t]
		\caption{Execution-Regulated Target-Field Reference}
		\label{alg:execution_regulated_reference}
		\begin{algorithmic}[1]
			\REQUIRE Ordered discrete path $Q^\star=(q_0,\ldots,q_T)$
			\ENSURE $\mathcal{T}_d(t)$, $\dot{\mathbf{c}}_d(t)$, and $\dot{\bar{\theta}}_d(t)$
			\STATE Initialize $l\leftarrow0$, $s\leftarrow0$, and $b_{\rm sw}\leftarrow0$
			\STATE Set $\mathcal{T}_d\leftarrow\mathcal{T}(q_0)$
			\WHILE{$l<T$}
			\IF{$k_{q_{l+1}}\ne k_{q_l}$}
			\STATE Hold the target-field pose and set $\dot{\mathbf{c}}_d\leftarrow\mathbf{0}$ and $\dot{\bar{\theta}}_d\leftarrow0$
			\IF{$b_{\rm sw}=0$}
			\STATE Evaluate the pre-switch guard in \eqref{eq:pre_switch_guard}
			\IF{\eqref{eq:pre_switch_guard} is satisfied}
			\STATE Update $k_d\leftarrow k_{q_{l+1}}$
			\STATE Set $t_{\rm sw}\leftarrow t$ and $b_{\rm sw}\leftarrow1$
			\ENDIF
			\ELSE
			\STATE Hold the post-switch target field $\mathcal{T}_d^+$
			\STATE Evaluate the post-switch release guard in \eqref{eq:post_switch_release_guard}
			\IF{\eqref{eq:post_switch_release_guard} is satisfied}
			\STATE Set $l\leftarrow l+1$, $s\leftarrow0$, and $b_{\rm sw}\leftarrow0$
			\STATE Set $\mathcal{T}_d\leftarrow\mathcal{T}(q_l)$
			\ENDIF
			\ENDIF
			\ELSE
			\STATE Construct $\mathcal{T}_d(t)$ using \eqref{eq:continuous_center_reference}
			\STATE Evaluate $M_{\rm ent}(\mathcal{T}_d)$, $M_{\rm cov}(\mathcal{T}_d)$, and $e_{\rm lag}(\mathcal{T}_d)$
			\STATE Compute $\phi(t)$ using \eqref{eq:execution_score}
			\STATE Compute $\alpha(t)$ using \eqref{eq:reference_speed_modulation}
			\STATE Update $s$ using \eqref{eq:reference_speed_modulation} and enforce $s\le1$
			\STATE Update $\mathbf{c}_d$ and $\theta_d$ using \eqref{eq:continuous_center_reference}
			\STATE Compute $\dot{\mathbf{c}}_d$ and $\dot{\bar{\theta}}_d$ using \eqref{eq:target_field_velocity}
			\IF{$s\ge1$}
			\STATE Set $\mathcal{T}_d\leftarrow\mathcal{T}(q_{l+1})$
			\STATE Set $l\leftarrow l+1$ and $s\leftarrow0$
			\ENDIF
			\ENDIF
			\ENDWHILE
			\STATE Hold $\mathcal{T}_d\leftarrow\mathcal{T}(q_T)$
			\STATE Set $\dot{\mathbf{c}}_d\leftarrow\mathbf{0}$ and $\dot{\bar{\theta}}_d\leftarrow0$
		\end{algorithmic}
	\end{algorithm}

	Algorithm~\ref{alg:execution_regulated_reference} combines both modes. The flag $b_{\rm sw}\in\{0,1\}$ records whether the active switching edge has updated its template. The manager changes timing and switch release without changing the planned spatial sequence. Its output is $\mathcal{T}_d(t)$ and the field velocities in \eqref{eq:target_field_velocity}. The next section converts these references into robot commands while retaining local assignment-free redistribution.

	\section{Morphology-Preserving Mean-Shift Navigation}
	\label{sec:mean_shift}
	The navigation law converts the regulated field reference into three robot-level contributions: reference motion, local redistribution, and obstacle correction. We first present the complete command and its moving-frame interpretation, then specify the local responses and execution corrections.

	\subsection{Control Structure and Moving-Frame Compensation}
	\label{subsec:dynamic_target_field_frame}
\begin{figure}[t]
		\centering
		\includegraphics[width=0.98\columnwidth,trim=0.2cm 0.2cm 0.2cm 0.2cm,clip]
		{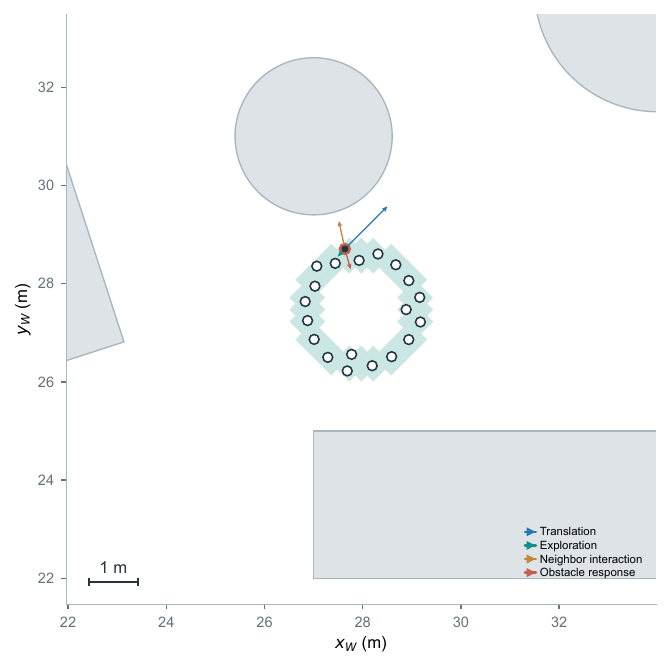}
		\caption{Instantaneous velocity-component decomposition for a representative robot. The arrows show target-field translation, assignment-free exploration, pairwise interaction, and obstacle response, all drawn from the robot center using a common display scale.}
		\label{fig:velocity_decomposition}
	\end{figure}

	The controller uses the field translation $\dot{\mathbf{c}}_d$ and continuous angular rate $\dot{\bar{\theta}}_d$ from Section~\ref{subsec:execution_regulated_field_reference}. With $J=[\begin{smallmatrix}0&-1\\1&0\end{smallmatrix}]$, its overall command is
	\begin{equation}
		\begin{aligned}
			\mathbf{u}_i&=\operatorname{sat}_{u_{\max}}(\mathbf{u}_i^{\rm nom}),\\
			\mathbf{u}_i^{\rm nom}&=\dot{\mathbf{c}}_d+\dot{\bar{\theta}}_dJ(\mathbf{p}_i-\mathbf{c}_d)\\
			&\quad+R(\theta_d)\mathbf{v}_i^L+\mathbf{u}_i^{\rm obs}.
		\end{aligned}
		\label{eq:overall_navigation_law}
	\end{equation}
	The first two terms execute the rigid field motion at the robot's current position. The residual $\mathbf{v}_i^L$ redistributes robots within the morphology, while $\mathbf{u}_i^{\rm obs}$ responds to nearby obstacles. Radial saturation enforces the speed limit. Sections~\ref{subsec:assignment_free_mean_shift_response} and \ref{subsec:morphology_preserving_navigation_law} define these components. Fig.~\ref{fig:velocity_decomposition} illustrates representative velocity contributions.

	To explain why this structure separates reference tracking from redistribution, express robot motion in the target-field frame. As in \eqref{eq:robot_local_coordinate}, $\mathbf{p}_i^L=R(\theta_d)^\top(\mathbf{p}_i-\mathbf{c}_d)$. On a non-switching interval, the continuous angle satisfies $\theta_d=\operatorname{wrap}(\bar{\theta}_d)$ and $R(\theta_d)=R(\bar{\theta}_d)$. Thus, $\dot{\bar{\theta}}_d$ represents physical angular velocity even across the wrapped-angle boundary. Differentiation gives
	\begin{equation}
		\dot{\mathbf{p}}_i^L = R(\theta_d)^\top \left( \mathbf{u}_i-\dot{\mathbf{c}}_d \right) - \dot{\bar{\theta}}_d J\mathbf{p}_i^L .
		\label{eq:dynamic_local_coordinate}
	\end{equation}
	The two frame-motion terms explain the compensation in \eqref{eq:overall_navigation_law}. Define the rigid velocity and nominal field-tracking component as
	\begin{equation}
		\begin{aligned}
			\mathbf{u}_i^{\rm rig}&=\dot{\mathbf{c}}_d+\dot{\bar{\theta}}_dJ(\mathbf{p}_i-\mathbf{c}_d),\\
			\mathbf{u}_i^{\rm tf}&=\mathbf{u}_i^{\rm rig}+R(\theta_d)\mathbf{v}_i^L.
		\end{aligned}
		\label{eq:target_field_velocity_decomposition}
	\end{equation}
	Translation is common to all robots, whereas rotation supplies a position-dependent tangential velocity. Neither component assigns a robot to a target cell. The resulting local motion is characterized by the following proposition.

	\begin{proposition}
		\label{prop:dynamic_frame_decoupling}
		Consider a non-switching interval on which $\mathbf{c}_d(t)$ and $\bar{\theta}_d(t)$ are differentiable. If robot $i$ applies the command in \eqref{eq:target_field_velocity_decomposition} without velocity saturation or additional robot-level correction, then
		\begin{equation}
			\dot{\mathbf{p}}_i^L = \mathbf{v}_i^L .
			\label{eq:decoupled_local_dynamics}
		\end{equation}
	\end{proposition}

	\begin{IEEEproof}
		Substituting the two relations in \eqref{eq:target_field_velocity_decomposition} into \eqref{eq:dynamic_local_coordinate} yields
		\begin{equation*}
			\begin{aligned}
				\dot{\mathbf{p}}_i^L ={}& R(\theta_d)^\top \left[ \dot{\bar{\theta}}_d J(\mathbf{p}_i-\mathbf{c}_d) \right]
				\\
				&+ R(\theta_d)^\top R(\theta_d) \mathbf{v}_i^L - \dot{\bar{\theta}}_d J\mathbf{p}_i^L .
			\end{aligned}
		\end{equation*}
		Using $\mathbf{p}_i-\mathbf{c}_d=R(\theta_d)\mathbf{p}_i^L$, $R(\theta_d)^\top R(\theta_d)=I$, and $R(\theta_d)^\top J R(\theta_d)=J$, the two rotational terms cancel, which gives \eqref{eq:decoupled_local_dynamics}.
	\end{IEEEproof}

	The proposition identifies a kinematic cancellation: after nominal frame compensation, local motion is governed by $\mathbf{v}_i^L$. It establishes neither redistribution convergence nor robot-level safety. Obstacle correction and saturation modify this relation as shown in Section~\ref{subsec:morphology_preserving_navigation_law}.

	\textit{Remark 4:} The decomposition in \eqref{eq:target_field_velocity_decomposition} should not be interpreted as the first mechanism for tracking a maneuvering desired shape. The method in \cite{sun2023meanshift} already propagates the translation and rotational motion of a desired shape through informed robots, distributed negotiation, and velocity alignment. In the present framework, the target-field motion is instead supplied directly by the environment-dependent planning and reference-management layers. Equation~\eqref{eq:target_field_velocity_decomposition} evaluates the complete rigid velocity at each robot position, including the position-dependent tangential component required by target-field rotation, and reserves the residual velocity for local assignment-free redistribution.

	\subsection{Assignment-Free Mean-Shift Response}
	\label{subsec:assignment_free_mean_shift_response}
	The local controller retains the entering, occupancy-redistribution, and neighbor-interaction mechanisms of assignment-free mean-shift exploration \cite{sun2023meanshift}. Their role in \eqref{eq:overall_navigation_law} is summarized first:
	\begin{equation}
		\mathbf{v}_i^L=\mathbf{w}_i^{{\rm in},L}+\mathbf{w}_i^{{\rm occ},L}+\mathbf{w}_i^{{\rm nbr},L}.
		\label{eq:assignment_free_mean_shift_response}
	\end{equation}
	All components use the active target-field frame. They regulate relative robot distribution; rigid field motion is supplied separately by \eqref{eq:target_field_velocity_decomposition}.

	\textit{1) Entering the target region:} Let $\mathcal{G}_k=\{(r,c)\mid r=0,\ldots,N_r^k-1,\ c=0,\ldots,N_c^k-1\}$ be the complete template grid, extending $\mathbf{x}_{k,\nu}^L$ to all cells. The target set remains $\mathcal{B}_k$, where $\Gamma_k=0$. The normalized guidance field increases with grid distance from the target region; its nonsaturated support is $\mathcal{D}_k=\{\nu\in\mathcal{G}_k\mid\Gamma_k(\nu)<1\}$. Let $\nu_i$ contain $\mathbf{p}_i^L$, or be the nearest valid cell outside the finite grid. Its adjacent-cell neighborhood is $\mathcal{H}_i=\{\nu\in\mathcal{G}_{k_d}\mid\|\nu-\nu_i\|_\infty\le1\}$. The entering cell is selected by
	\begin{equation}
		\nu_i^{\rm in}\in
		\begin{cases}
			\displaystyle\mathop{\arg\min}_{\nu\in\mathcal{D}_{k_d}}\left\|\mathbf{x}_{k_d,\nu}^L-\mathbf{p}_i^L\right\|, & \nu_i\in\mathcal{G}_{k_d}\setminus\mathcal{D}_{k_d},\\[3mm]
			\displaystyle\mathop{\arg\min}_{\nu\in\mathcal{H}_i}\Gamma_{k_d}(\nu), & \nu_i\in\mathcal{D}_{k_d}\setminus\mathcal{B}_{k_d},\\[3mm]
			\{\nu_i\}, & \nu_i\in\mathcal{B}_{k_d}.
		\end{cases}
		\label{eq:entering_descent_cell}
	\end{equation}
	Ties are resolved by the closest cell center. The first case reaches the guidance support; the second descends toward the target region. This selection produces
	\begin{equation}
		\mathbf{w}_i^{{\rm in},L}=a_{\rm in}\Gamma_{k_d}(\nu_i)\frac{\mathbf{x}_{k_d,\nu_i^{\rm in}}^L-\mathbf{p}_i^L}{\left\|\mathbf{x}_{k_d,\nu_i^{\rm in}}^L-\mathbf{p}_i^L\right\|+\varepsilon_d},
		\label{eq:local_entering_response}
	\end{equation}
	Here, $a_{\rm in}>0$ and $\varepsilon_d>0$ is a small distance regularizer. The response vanishes on target cells because $\Gamma_{k_d}=0$, leaving redistribution and interactions to organize robots inside the shape.

	\textit{2) Exploring available target cells:} Exploration directs robots toward a local density of admissible target cells. With sensing radius $r_{\rm sense}>0$, neighbors satisfy $\mathcal{N}_i=\{j\ne i\mid\|\mathbf{p}_j-\mathbf{p}_i\|\le r_{\rm sense}\}$. For every $\nu\in\mathcal{G}_{k_d}$, let $d_{i,\nu}=\|\mathbf{p}_i^L-\mathbf{x}_{k_d,\nu}^L\|$. Local occupancy, visible targets, and the boundary indicator are collected as
	\begin{equation}
		\begin{aligned}
			z_{i,\nu}&=\mathbb{I}\!\left[\min_{j\in\mathcal{N}_i\cup\{i\}}\|\mathbf{p}_j^L-\mathbf{x}_{k_d,\nu}^L\|\le d_{\rm occ}\right],\\
			\mathcal{C}_i&=\{\nu\in\mathcal{B}_{k_d}\mid d_{i,\nu}\le r_{\rm sense}\},\\
			b_i&=\mathbb{I}[\exists\nu\in\mathcal{G}_{k_d}\setminus\mathcal{B}_{k_d},\ d_{i,\nu}\le r_{\rm sense}].
		\end{aligned}
		\label{eq:local_cell_occupancy}
	\end{equation}
	For hollow templates such as rings, boundary mode is disabled by setting $b_i=0$. Otherwise, all visible target cells contribute near a boundary, while occupied cells are excluded in the interior. Hence,
	\begin{equation}
		\mathcal{A}_i=
		\begin{cases}
			\mathcal{C}_i, & b_i=1,\\[1mm]
			\left\{\nu\in\mathcal{C}_i\,\middle|\,z_{i,\nu}=0\right\}, & b_i=0.
		\end{cases}
		\label{eq:valid_mean_shift_cell_set}
	\end{equation}
	Use the compactly supported kernel
	\begin{equation*}
		K_{\rm ms}(s)=\begin{cases}
			1,&s\le0,\\
			\tfrac12[1+\cos(\pi s)],&0<s<1,\\
			0,&s\ge1.
		\end{cases}
	\end{equation*}
	With a dimensionless regularizer $\varepsilon_w>0$, the displacement and occupancy response are
	\begin{equation}
		\begin{aligned}
			\mathbf{h}_i^L&=\frac{\displaystyle\sum_{\nu\in\mathcal{A}_i}K_{\rm ms}\!\left(\frac{d_{i,\nu}}{r_{\rm sense}}\right)(\mathbf{x}_{k_d,\nu}^L-\mathbf{p}_i^L)}{\displaystyle\sum_{\nu\in\mathcal{A}_i}K_{\rm ms}\!\left(\frac{d_{i,\nu}}{r_{\rm sense}}\right)+\varepsilon_w},\\
			\mathbf{w}_i^{{\rm occ},L}&=a_i^{\rm ms}\mathbf{h}_i^L,\qquad
			a_i^{\rm ms}=\begin{cases}a_{\rm bd},&b_i=1,\\a_{\rm fill},&b_i=0.
			\end{cases}
		\end{aligned}
		\label{eq:local_exploration_response}
	\end{equation}
	The positive gains $a_{\rm bd}$ and $a_{\rm fill}$ tune boundary entry and interior filling, respectively. If $\mathcal{A}_i$ is empty, the response is zero. Both modes redistribute robots without reserving a cell for any individual.

	\textit{3) Neighbor interactions:} Let $d_{ij}=\|\mathbf{p}_i-\mathbf{p}_j\|=\|\mathbf{p}_i^L-\mathbf{p}_j^L\|$. For distinct centers, separation is activated below the desired distance $d_{\rm r}$:
	\begin{equation}
		\begin{aligned}
			\mathbf{w}_i^{{\rm sep},L}&=a_{\rm sep}\sum_{j\in\mathcal{N}_i}\lambda_{\rm sep}(d_{ij})\frac{\mathbf{p}_i^L-\mathbf{p}_j^L}{d_{ij}+\varepsilon_d},\\
			\lambda_{\rm sep}(d)&=\begin{cases}d_{\rm r}/(d+\varepsilon_d)-1,&0<d<d_{\rm r},\\0,&d\ge d_{\rm r}.
			\end{cases}
		\end{aligned}
		\label{eq:local_robot_repulsion_response}
	\end{equation}
	Here, $a_{\rm sep}>0$. Velocity alignment complements separation. Let $\mathbf{u}_i^-$ denote the world-frame velocity command from the preceding control update. Expressing its neighbor differences in the current target-field frame gives
	\begin{equation}
		\begin{aligned}
			\mathbf{w}_i^{{\rm ali},L}&=\frac{a_{\rm ali}}{|\mathcal{N}_i|+\varepsilon_w}R(\theta_d)^\top\!\sum_{j\in\mathcal{N}_i}(\mathbf{u}_j^--\mathbf{u}_i^-),\\
			\mathbf{w}_i^{{\rm nbr},L}&=\mathbf{w}_i^{{\rm sep},L}+\mathbf{w}_i^{{\rm ali},L},
		\end{aligned}
		\label{eq:local_interaction_response}
	\end{equation}
	where $a_{\rm ali}\ge0$. Alignment uses total preceding commands, including their reference-motion components; it does not subtract rigid field motion. The prescribed translation and tangential velocities are supplied explicitly by $\mathbf{u}_i^{\rm rig}$. When $k_d$ changes, the same responses are recomputed using the new target cells and guidance field.

	\subsection{Execution Corrections and Closed-Loop Kinematics}
	\label{subsec:morphology_preserving_navigation_law}
	Two execution effects modify the nominal decomposition: local obstacle responses and velocity saturation. This subsection specifies both and shows how they enter the local kinematics, completing the command in \eqref{eq:overall_navigation_law}.

	\textit{1) Local obstacle correction:} Robots may leave the target region during entry or redistribution, so field feasibility alone does not ensure robot-level clearance. Let $\mathcal{I}_i^{\rm o}\subseteq\{1,\ldots,N_o\}$ index locally detected obstacles. For each $m$, select $\mathbf{q}_{im}\in\arg\min_{\mathbf{q}\in O_m}\|\mathbf{p}_i-\mathbf{q}\|$ and set $d_{im}=\|\mathbf{p}_i-\mathbf{q}_{im}\|$. This is the center-to-obstacle distance used in Section~\ref{sec:problem_formulation}. The outward direction and activation are
	\begin{equation}
		\begin{aligned}
			\mathbf{n}_{im}&=\frac{\mathbf{p}_i-\mathbf{q}_{im}}{d_{im}+\varepsilon_d},\\
			\beta_{\rm o}(d)&=\max\left\{0,\frac{d_{\rm inf}}{d+\varepsilon_d}-1\right\},\quad d_{\rm inf}>d_{\rm o}.
		\end{aligned}
		\label{eq:obstacle_activation_function}
	\end{equation}
	The influence distance $d_{\rm inf}$ is the robot radius plus an obstacle-response margin. The response is inactive beyond this distance and increases as the robot approaches the obstacle. Summing locally detected contributions gives
	\begin{equation}
		\mathbf{u}_i^{\rm obs}=a_{\rm obs}\sum_{m\in\mathcal{I}_i^{\rm o}}\beta_{\rm o}(d_{im})\mathbf{n}_{im},
		\label{eq:local_obstacle_response}
	\end{equation}
	The gain $a_{\rm obs}>0$ scales this world-frame correction. It reduces obstacle proximity but does not formally enforce \eqref{eq:robot_obstacle_clearance} at every instant.

	\textit{2) Velocity limits and closed-loop behavior:} Radial saturation preserves command direction and enforces the speed bound:
	\begin{equation}
		\operatorname{sat}_{u_{\max}}(\mathbf{v})=
		\frac{u_{\max}}{\max\{u_{\max},\|\mathbf{v}\|\}}\mathbf{v}.
		\label{eq:radial_velocity_saturation}
	\end{equation}
	Define $\boldsymbol{\delta}_i=\mathbf{u}_i-\mathbf{u}_i^{\rm nom}$, which vanishes when saturation is inactive. Substituting the command in \eqref{eq:overall_navigation_law} into \eqref{eq:dynamic_local_coordinate} gives
	\begin{equation}
		\dot{\mathbf{p}}_i^L=\mathbf{v}_i^L+R(\theta_d)^\top\mathbf{u}_i^{\rm obs}+R(\theta_d)^\top\boldsymbol{\delta}_i.
		\label{eq:closed_loop_local_robot_dynamics}
	\end{equation}
	To relate reference progression to the available robot speed, consider a differentiable instant of an active non-switching edge. Define the rigid-motion demand bound $A_i(t)$ and the local-response bound $B_i(t)$ as
	\begin{equation}
		\begin{aligned}
			A_i(t)&=\frac{\|\Delta\mathbf{c}_l\|+
				|\Delta\theta_l|\,\|\mathbf{p}_i-\mathbf{c}_d\|}{\tau_l},\\
			B_i(t)&=\|\mathbf{v}_i^L\|+\|\mathbf{u}_i^{\rm obs}\|.
		\end{aligned}
		\label{eq:execution_velocity_budget}
	\end{equation}
	Both bounds have units of speed and use the current robot state and local responses.

	\begin{proposition}[Velocity feasibility and local-motion deviation]
		\label{prop:velocity_feasibility_residual_bound}
		For the controller in \eqref{eq:overall_navigation_law}, robot $i$ executes its nominal velocity without saturation whenever
		\begin{equation}
			\alpha(t)A_i(t)+B_i(t)\le u_{\max}.
			\label{eq:unsaturated_execution_condition}
		\end{equation}
		At every differentiable instant of a non-switching segment, with or without saturation, the local-motion deviation satisfies
		\begin{equation}
			\begin{aligned}
				\|\dot{\mathbf{p}}_i^L-\mathbf{v}_i^L\|
				&\le\|\mathbf{u}_i^{\rm obs}\|\\
				&\quad+\bigl(\|\mathbf{u}_i^{\rm nom}\|-u_{\max}\bigr)_+,
			\end{aligned}
			\label{eq:local_motion_deviation_bound}
		\end{equation}
		where $(z)_+=\max\{z,0\}$. If \eqref{eq:unsaturated_execution_condition} holds and $\mathbf{u}_i^{\rm obs}=\mathbf{0}$, then $\dot{\mathbf{p}}_i^L=\mathbf{v}_i^L$.
	\end{proposition}

	\begin{IEEEproof}
		Using the field velocities in \eqref{eq:target_field_velocity}, $\|J\mathbf{x}\|=\|\mathbf{x}\|$, and rotation invariance of the Euclidean norm gives $\|\mathbf{u}_i^{\rm nom}\|\le\alpha A_i+B_i$, since $\alpha\ge0$. Thus, \eqref{eq:unsaturated_execution_condition} implies that radial saturation is inactive. From \eqref{eq:radial_velocity_saturation}, the saturation residual has the exact magnitude $\|\boldsymbol{\delta}_i\|=(\|\mathbf{u}_i^{\rm nom}\|-u_{\max})_+$. Applying the triangle inequality to \eqref{eq:closed_loop_local_robot_dynamics} yields \eqref{eq:local_motion_deviation_bound}. When both corrections vanish, the same equation reduces to \eqref{eq:decoupled_local_dynamics}.
	\end{IEEEproof}

	The sufficient condition reserves speed for redistribution and obstacle response after accounting for rigid reference motion. Rotation demands more speed from robots farther from the field center. For fixed robot states and local responses, reducing $\alpha$ reduces the rigid-motion bound $\alpha A_i$; if $B_i>u_{\max}$, however, this sufficient condition cannot be met by slowing the reference alone. The metric-based manager does not explicitly enforce \eqref{eq:unsaturated_execution_condition}, which is a pointwise execution check rather than an additional control rule. The deviation bound quantifies obstacle and saturation effects without asserting shape convergence or robot-level safety.

	During a switch or hold, $\dot{\mathbf{c}}_d=\mathbf{0}$ and $\dot{\bar{\theta}}_d=0$, so $\mathbf{u}_i^{\rm rig}=\mathbf{0}$ and $\mathbf{u}_i^{\rm nom}=R(\theta_d)\mathbf{v}_i^L+\mathbf{u}_i^{\rm obs}$. Robots then reorganize relative to the held field; motion after switching resumes according to \eqref{eq:post_switch_release_guard}.

	Here, ``morphology-preserving'' refers to coherent occupation of the evolving unlabeled target field. Propositions~\ref{prop:dynamic_frame_decoupling} and \ref{prop:velocity_feasibility_residual_bound} characterize nominal frame compensation and execution deviations, respectively. Section~\ref{sec:experiments} evaluates navigation with obstacle responses, switching, and velocity limits.

	\section{Simulation and Experimental Results}
	\label{sec:experiments}
	The evaluation examines whether an adaptive target field can guide assignment-free swarm navigation across different populations and obstacle environments. Numerical simulations assess population applicability, joint position--orientation--morphology planning, and the role of morphology selection. Hardware experiments assess execution on physical robots under motion-capture feedback.

	\subsection{Numerical Simulations}
	\label{subsec:numerical_simulations}
	All simulations used disk robots of radius $0.10$~m and an integration step of $0.05$~s. Target-field routes were planned before execution and then advanced using entering, coverage, and centroid-lag feedback, without online replanning. Terminal quality is reported using the entering and coverage metrics defined in Section~\ref{sec:problem_formulation}; the superscript $g$ denotes the terminal state.
	Table~\ref{tab:execution_parameters} lists key settings using archived gain names.

	\subsubsection{Navigation across swarm populations}
	\label{subsec:population_scalability}
	To assess whether the same target-field representation could guide different swarm populations, we simulated $N\in\{10,20,50,100\}$ robots in geometrically similar two-gate environments. The candidate library comprised a circle ring and horizontal and vertical line templates, all admitting explicit rotation. The workspace, template dimensions, and planning grid were scaled by
	\begin{equation}
		s_N=\sqrt{\frac{N}{10}}.
		\label{eq:swarm_scale_factor}
	\end{equation}
	The baseline workspace was $15\times10$~m. The reference speed was $0.35s_N$~m/s, while the robot speed limit remained $1.50$~m/s. The interaction gain followed $\kappa_3=0.25\max\{1,N/50\}$. Thus, the study tested navigation across populations with geometric and speed scaling, rather than computational complexity at fixed settings.
	The same $64\times64$ ring (1,121 target cells) and $24\times65$ line (392 target cells) were reused across populations, with physical cell lengths set by \eqref{eq:cell_length}. Although rotation was available, the two-gate routes used the horizontal and vertical line candidates without rotation actions. Explicit reorientation was evaluated separately in the heterogeneous scene.

\begin{figure*}[!t]
			\centering
			\includegraphics[width=0.90\textwidth,trim=0 20bp 0 3bp,clip]{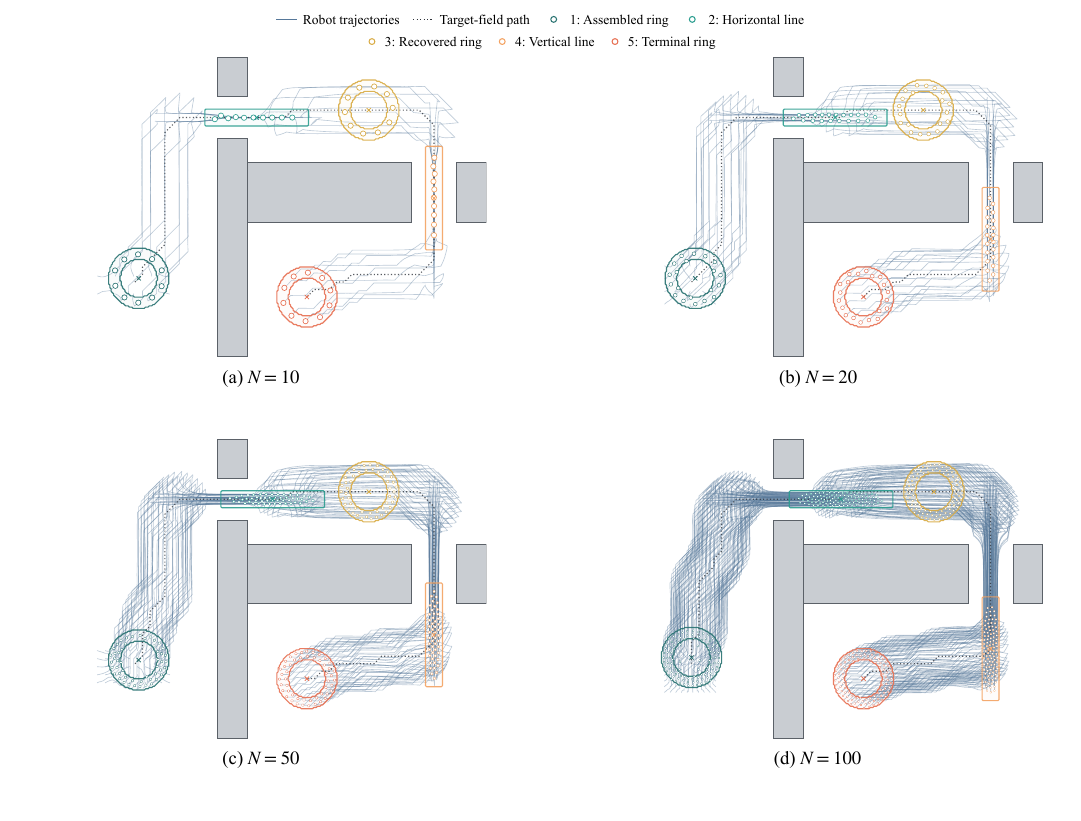}
			\caption{Navigation with $N=10$, 20, 50, and 100 robots in geometrically similar two-gate environments. Curves show recorded trajectories; colors identify assembly, line passages, ring recovery, and arrival. Crosses and the dotted curve indicate field centers and the planned path. Robot radius is $0.10$~m; environments scale according to \eqref{eq:swarm_scale_factor}.}
			\label{fig:scalability}
	\end{figure*}

	As shown in Fig.~\ref{fig:scalability}, every population passed both gates by adopting the corresponding line morphology and recovering the ring afterward. The planner selected these changes without prescribed switching locations, and the robots realized them through local redistribution without persistent robot-to-cell assignments. All runs reached $100\%$ terminal entering, with coverage exceeding $82\%$ (Table~\ref{tab:scalability}); no contact event was recorded. These results support the use of the same planning-to-execution representation across the tested populations under the stated scaling protocol.

	\begin{table}[H]
		\caption{Terminal navigation results across swarm populations.}
		\label{tab:scalability}
		\centering
		\footnotesize
		\setlength{\tabcolsep}{3.0pt}
		\renewcommand{\arraystretch}{1.10}
		\begin{tabular*}{\columnwidth}{@{\extracolsep{\fill}}rrrr@{}}
			\toprule
			$N$ & $T_{\rm goal}$ (s) & $M_{\rm ent}^{g}$ (\%) & $M_{\rm cov}^{g}$ (\%) \\
			\midrule
			10 & 82.50 & 100 & 87.33 \\
			20 & 81.15 & 100 & 82.60 \\
			50 & 84.20 & 100 & 88.67 \\
			100 & 89.65 & 100 & 89.21 \\
			\bottomrule
		\end{tabular*}
	\end{table}
	Total post-switch holding ranged from $5.10$ to $12.70$~s across the four runs, with the longest hold recorded for $N=100$. During these intervals, the reference pose remained fixed while robots redistributed within the updated field. Thus, the common morphology sequence was realized with population-dependent transition timing rather than a prescribed switching timetable.

	\begin{figure}[H]
		\centering
		\includegraphics[width=1.02\columnwidth]{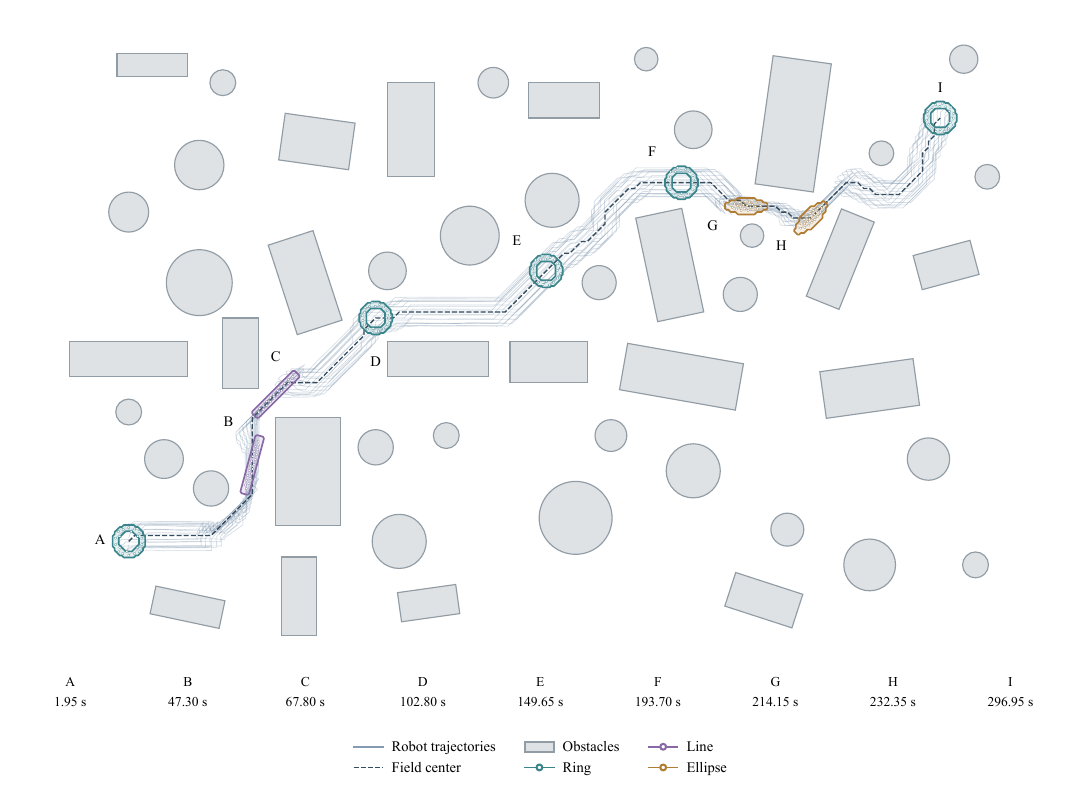}
		\caption{Navigation in the heterogeneous obstacle environment. Pale blue curves show robot trajectories, the dashed curve shows the target-field center path, and gray regions denote obstacles. Recorded configurations show ring assembly (A), line passage and reorientation (B--C), ring recovery (D--F), ellipse passage and reorientation (G--H), and terminal ring recovery (I). Times are given below the map. Robot disks use the model radius of $0.10$~m.}
		\label{fig:exp2_complex_navigation}
	\end{figure}
	\begin{figure*}[!t]
		\centering
		\includegraphics[width=0.94\textwidth]{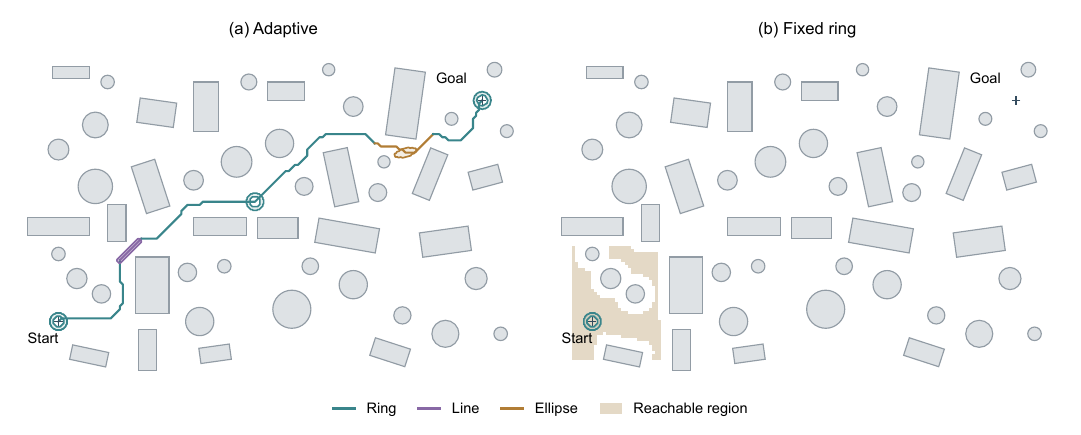}
		\caption{Planning with and without morphology selection; both variants allow translation and rotation. (a) Path colors indicate the active morphology, and outlines show representative planned fields. (b) Shading shows reachable field-center locations for the fixed ring, whose initial footprint is drawn at the same spatial scale. The goal lies outside the start component of the tested discrete graph.}
		\label{fig:exp2_fixed_ring_ablation}
	\end{figure*}
	\subsubsection{Adaptation in a heterogeneous environment}
	\label{subsec:exp2_complex_environment}
	We next tested joint position, orientation, and morphology adaptation with 20 robots in an $80\times52$~m workspace containing 29 circular and 20 rectangular obstacles. The library contained 12 templates, with a circle ring preferred at the start and goal. The planner used a $0.50$~m position grid and $15^{\circ}$ heading increments; no intermediate waypoint or morphology sequence was supplied. The reference and robot speed limits were $0.35$ and $1.0$~m/s, respectively.

	The selected route combined translation and rotation with the sequence \emph{circle ring $\rightarrow$ line $\rightarrow$ circle ring $\rightarrow$ ellipse $\rightarrow$ circle ring} (Fig.~\ref{fig:exp2_complex_navigation}). The line reoriented through the lower bottleneck, while the ellipse negotiated the upper restricted region. The swarm then recovered the terminal ring with $100\%$ entering and $83.33\%$ coverage at $296.95$~s. This behavior illustrates how planning the collective field provides environment-dependent guidance without specifying individual robot goals. Relative to the underlying mean-shift shape-assembly and maneuvering controller \cite{sun2023meanshift}, the added capability is environment-dependent route, orientation, and morphology selection, coupled with feedback-regulated progression and switching.

	During morphology changes, the reference manager held the updated field while the robots redistributed. Between switches, the nominal run required no reduction of reference speed. Local redistribution also produced transiently nonuniform occupation: full entering in the ellipse configurations did not imply full coverage. No contact was recorded in this nominal run, although the minimum robot--robot surface gap was only $0.0154$~m. These observations demonstrate complete-framework execution in the scene without establishing continuous robot-level safety.
	The transition records clarify this distinction. Immediately after the first ring-to-line update, entering and coverage relative to the new field dropped to $20\%$ and $10\%$, respectively. The switch guards held the field for $4.90$~s in total, including $0.90$~s beyond the four minimum dwell intervals. In the later ellipse configurations (G--H), entering reached $100\%$ while coverage remained $65.3\%$. These observations distinguish reaching the target region from distributing throughout it, motivating their separate use in execution assessment.

	\subsubsection{Role of morphology selection}
	To isolate morphology selection, a paired test compared the full library with a library restricted to the circle ring, keeping the environment, discretization, shared action costs, and execution parameters unchanged. Direct ring rotation, disabled in the preceding nominal run, was enabled in both variants. The adaptive variant was therefore replanned and executed separately.
	Across the archived runs, translation/rotation/switch coefficients were $1/0.10/1.0$, and template-use costs were $0/0.20$ for preferred/other morphologies. The clearance weight in \eqref{eq:transition_cost_components} was $w_{\rm clr}=3$. Scene-specific margins and $d_{\rm pref}$ are supplied in the configuration archive.

	\begin{table}[!t]
		\caption{Key execution settings.}
		\label{tab:execution_parameters}
		\centering
		\footnotesize
		\setlength{\tabcolsep}{3.0pt}
		\renewcommand{\arraystretch}{1.05}
		\begin{tabular}{@{}p{0.37\columnwidth}p{0.27\columnwidth}p{0.29\columnwidth}@{}}
			\toprule
			Parameter & Simulation & Hardware \\
			\midrule
			$d_{\rm in},d_{\rm occ}$ (m) & $0.25,0.25$ & $0.25,0.25$ \\
			$r_{\rm sense}$ (m) & $1.0$ & $0.50$ \\
			$\kappa_1;\sigma_1,\sigma_2$ & $0.25;2,1$ & $0.25;2,1$ \\
			$\kappa_3$ & $0.25$ & $0.25$ \\
			Obstacle-response gain & $0.80$ & $0.08$ \\
			Entry/coverage/lag gates & $60\%,55\%,0.25$~m & $60\%,55\%,0.25$~m \\
			$\phi_{\rm hold},\phi_{\rm nom};\alpha_{\rm slow}$ & $0.60,1.0;0.30$ & $0.60,1.0;0.30$ \\
			Min. post-switch (s) & $1.0$ & $1.0$ \\
			Update period (s) & $0.05$ & $0.05$ nominal \\
			Terminal entry threshold & $100\%$ & $90\%$ \\
			\bottomrule
		\end{tabular}
		\par\vspace{2pt}\raggedright
		Hardware mean measured update period: 61.76 ms. Gains $\kappa_1$, $\sigma_1/\sigma_2$, and $\kappa_3$ scale entering, boundary/interior exploration, and neighbor interactions, respectively. For simulations with $N=100$, $\kappa_3$ was increased to $0.50$; otherwise, $\kappa_3=0.25$.
	\end{table}

	The fixed-ring search exhausted the start component without finding a route, whereas the adaptive planner connected the same start and goal (Fig.~\ref{fig:exp2_fixed_ring_ablation}). Morphology selection therefore changed planning feasibility in the tested discrete graph. The new adaptive execution reached $100\%$ entering and $80.30\%$ coverage. The comparison supports the role of morphology adaptation in route feasibility; it does not imply collision-free robot execution.

	\begin{figure*}[!t]
		\centering
		\includegraphics[width=0.88\textwidth]{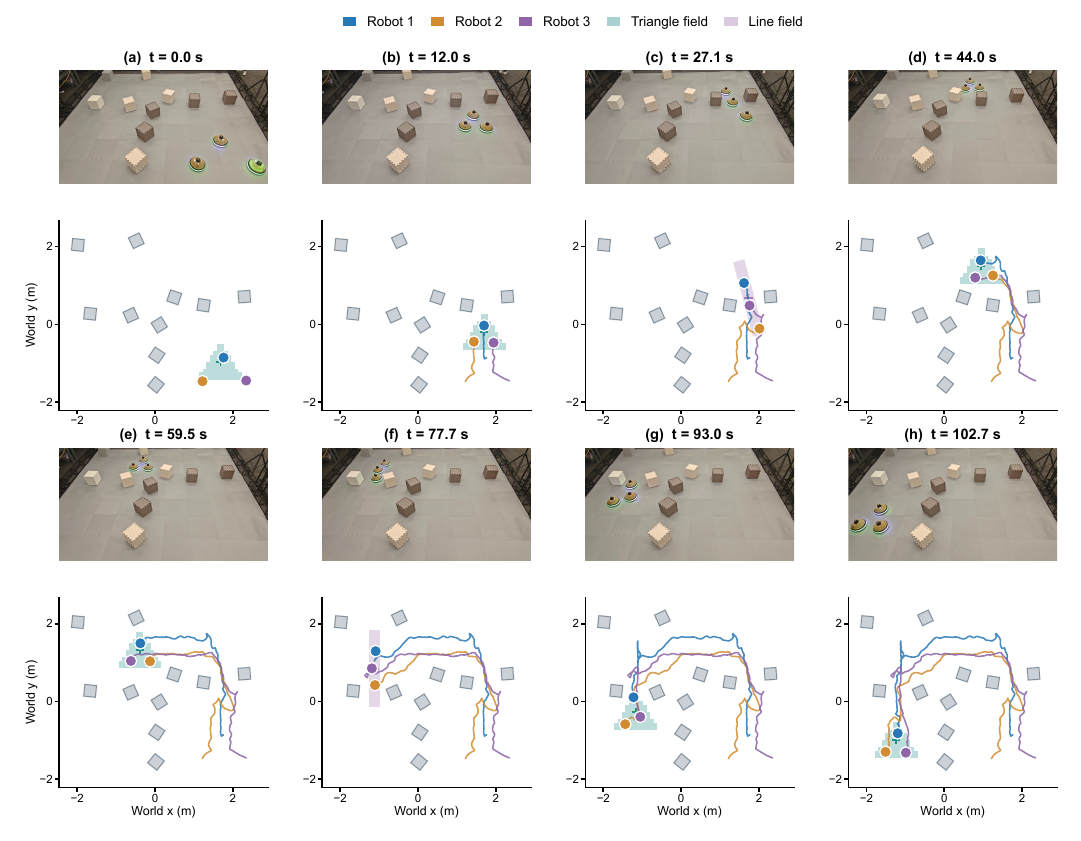}
		\caption{Hardware navigation with repeated morphology adaptation. Each group (a)--(h) pairs a video frame with a reconstruction from the same motion-capture logs at one shared elapsed time. Colored disks and curves show measured robot positions and cumulative trajectories; teal and purple regions denote triangle and line target fields. The sequence shows initial triangle motion (a, b), the first line passage (c), triangle recovery and translation (d, e), the second line passage (f), and final recovery and arrival (g, h). Labels 1--3 identify robots, without implying persistent target assignments.}
		\label{fig:hardware_navigation}
	\end{figure*}
    
	\subsection{Hardware Experiments}
	\label{subsec:hardware_validation}
	Hardware validation used three Mecanum-wheel robots of model radius $0.15$~m in an approximately $5.0\times4.5$~m workspace with ten square obstacles of side length $0.30$~m. A 12-camera NOKOV motion-capture system supplied poses to a central control computer, which computed world-frame commands and transmitted the corresponding body-frame velocities wirelessly. The planner used measured obstacle models with a $0.025$~m margin, a preferred triangle morphology, and rotatable line templates. Only planar translational commands were issued to the robots. Before transmission, a hardware protection layer adjusted the nominal commands using speed and acceleration limits and distance-rate constraints for inter-robot separation, workspace boundaries, and obstacles; this implementation layer is additional to the nominal controller analyzed above.

	As shown in Fig.~\ref{fig:hardware_navigation}, the robots completed two triangle--line--triangle passage-and-recovery cycles along the saved route. The narrow line field guided passage through constrained regions, and the preferred triangle was recovered in open space. After each switch, the reference manager held the field until the redistribution conditions were satisfied, then resumed progression. The measured sequence demonstrates physical execution of the adaptive collective reference without persistent robot-to-cell assignments.
	The feedback logs distinguish motion regulation from switch coordination. During non-switching motion, $\alpha$ decreased to $0.3$ for nine control frames, totaling approximately $0.56$~s; no motion-stage feedback pause was recorded. Post-switch settling occupied approximately $16.54$~s, exceeding the four minimum 1~s dwell intervals and showing that release also depended on the measured redistribution state. Thus, feedback intervention in this run was dominated by morphology-transition coordination. These records establish that the mechanisms were active, without isolating their performance benefit against a feedback-disabled baseline.

	The swarm reached the stable terminal condition at $102.67$~s with $100\%$ entering and $81.43\%$ coverage. The minimum sampled robot--robot and robot--obstacle-model surface gaps were $0.0947$ and $0.0677$~m, respectively. Thus, all robots entered the terminal field and maintained positive measured clearances in this hardware run, supporting the feasibility of the proposed planning-to-execution framework on physical robots. The clearances are sampled geometric observations rather than a continuous safety guarantee.
	The reference itself arrived at $101.68$~s, before the terminal stability check completed. Full entering together with partial coverage is consistent with the region-based objective: all three robots occupied the terminal field, while some target cells remained uncovered. The terminal centroid lag was $0.1455$~m.

	Completion tests also required coverage $\ge75\%$, centroid lag $\le0.25$~m, and mean target-cell distance $\le0.05$~m for 1~s after reference arrival. The distance diagnostic assigns zero inside target cells and does not enter $\phi$. The hardware entry threshold was $90\%$; the observed result was $100\%$.

	\section{Conclusion}
	\label{sec:conclusion}
	This paper presented an environment-adaptive planning-to-execution framework that extends assignment-free shape assembly to obstacle-constrained swarm navigation. Rather than specifying robot-specific goals, the framework plans the position, orientation, and morphology of an unlabeled target field. Feasibility-aware graph search determines its spatial evolution, an execution-regulated reference manager coordinates progression and morphology switching with the measured swarm state, and a moving-frame mean-shift controller drives local redistribution while accounting for field motion and nearby obstacles.

	The experiments support complementary aspects of this design. Geometrically scaled simulations with 10--100 robots demonstrated navigation and morphology recovery across the tested populations. In a heterogeneous environment, joint translation, rotation, and morphology selection produced a traversable route, while a fixed-ring ablation showed the role of morphology adaptation in planning feasibility. The motion-capture experiment further demonstrated two triangle--line--triangle passage-and-recovery cycles with three Mecanum robots among ten physical obstacles, reaching the terminal field with full entry and positive sampled geometric clearances. Together, these results support the use of an adaptive collective reference to connect environment-level planning with assignment-free local robot motion in the tested settings.

	Theoretical robot-level safety guarantee is an important direction for further extending the proposed framework. Although target-field feasibility supports collective navigation, guaranteeing inter-robot and obstacle clearance throughout execution requires additional safety mechanisms. Future work will investigate the integration of control barrier functions\cite{zhang2025conformal} to address these constraints while preserving assignment-free redistribution during target-field motion and morphology transitions.

	\bibliographystyle{IEEEtran}
	\bibliography{references_swarm_navigation_updated}

\end{document}